\documentclass[graybox,natbib]{svmult}

\usepackage{mathptmx}       
\usepackage{helvet}         
\usepackage{courier}        
\usepackage{type1cm}        
\usepackage{makeidx}         
\usepackage{graphicx}        
\usepackage{multicol}        
\usepackage[bottom]{footmisc}

\makeindex             

\newcommand{\yoeff}{$y_{\!\mbox{\scriptsize O,eff}}$}

\newcommand{\hi}{H\,{\sc i}\rm}
\newcommand{\hii}{H\,{\sc ii}\rm}

\newcommand{\nii}{[N\,{\sc ii}]}

\newcommand{\oiii}{[O\,{\sc iii}]}
\newcommand{\oii}{[O\,{\sc ii}]}
\newcommand{\oiir}{O\,{\sc ii}}

\newcommand{\te}{$T_{\rm e}$}

\newcommand{\hbeta}{H$\beta$}
\newcommand{\halpha}{H$\alpha$}
\newcommand{\lin}{$\,\lambda$}
\newcommand{\llin}{$\,\lambda\lambda$}

\newcommand{\rtf}{$R_{25}$}
\newcommand{\re}{$R_{\mathrm e}$}

\newcommand{\oh}{12\,+\,log(O/H)}

\newcommand{\ohsun}{\mbox{12\,+\,log(O/H)$_\odot\,=\,$}}
\newcommand{\rtwothree}{R23}

\newcommand{\eg}{e.g.}
\newcommand{\ie}{i.e.}

\begin{document}

\title*{Metallicities in the Outer Regions of Spiral Galaxies}
\author{Fabio Bresolin}
\institute{Fabio Bresolin \at Institute for Astronomy, 2680 Woodlawn Drive, Honolulu HI 96822, USA; \\
\email{bresolin@ifa.hawaii.edu}
}
\maketitle

\abstract{The analysis of the chemical composition of galaxies provides fundamental insights into their evolution. This holds true also in the case of the outer regions of spiral galaxies.
This Chapter presents the observational data, accumulated in the past few years mostly from the analysis of \hii\ region spectra, concerning the metallicity of the outer disks of spirals that are characterized by extended \hi\ envelopes and low star formation rates. 
I present evidence from the literature that the metal radial distribution flattens at large galactocentric distances, with levels of enrichment that exceed those expected given the large gas mass fractions and the weak star formation activity.
The interpretation of these results leads to speculations regarding mechanisms of metal mixing in galactic disks and the possibility that metal-enriched gas infall plays a role in determining the chemical evolution of the outskirts of spirals.
}

\section{Introduction}\label{sec:1}
The analysis of the chemical abundance composition of galaxies provides essential and unique constraints on their evolutionary status and their star formation properties. Gathering spatially resolved information about the distribution of 
metals is a well-tested approach to probe not only the metal production in stars across time, but also those effects, such as galactic wind outflows, gravitational interactions, secular processes and gas inflows, that can profoundly affect the evolution of galaxies.

This Chapter looks at the present-day gas metallicities\index{metallicities} of outer spiral disks, as derived from the emission line analysis of \hii\ region spectra, excluding older chemical abundance tracers, such as planetary nebulae and stars, except for a few notable exceptions. The connections between the chemical abundances of the outer disks thus derived and of the circumgalactic medium, probed by resonance absorption lines in the UV, for example in damped Lyman $\alpha$ systems, is covered elsewhere in this volume (see the review by Chen, this volume).

A non-secondary aspect of chemical abundance work in nearby and far-away star-forming galaxies concerns the methodology employed in the measurement of nebular (ionized gas) abundances. Therefore, Sect.~\ref{Sec:diagnostics} provides a brief overview of the difficulties and the techniques used. The subsequent sections provide details on the work carried out in a variety of nearby systems (Sect.~\ref{Sect:work} and \ref{Sect:add}), building the framework for interpreting the observed chemical abundance properties (Sect.~\ref{Sect:models}). A concise summary concludes the Chapter.

\section{Measuring Nebular Abundances}\label{Sec:diagnostics}

The measurement of nebular chemical abundances that are free of significant systematic uncertainties remains an unsolved problem in astrophysics, despite decades of observational efforts to investigate the emission-line properties of Galactic and extragalactic \hii\ regions.
Many authors (among others: \citealt*{Bresolin:2004,Kewley:2008}; \citealt{Lopez-Sanchez:2012}) have addressed this issue, showing how the various 
emission line diagnostics and the different calibrations proposed in the literature for these diagnostics are afflicted by systematic variations on the derived oxygen abundances,
that reach up to 0.7\,dex.\footnote{In the literature measuring the {\em metallicity} of an \hii\ region is equivalent to measuring its {\em oxygen abundance}\index{oxygen abundance} O/H, since O constitutes (by number) approximately half of the metals. The standard practice is to report the value of \oh. The Solar value used for reference is taken here to be \ohsun\,8.69, from \citet{Asplund:2009}.}
This problem, of course, affects not only the investigation of ionized nebulae in the local Universe, \eg, for the analysis of abundance gradients in spiral galaxies (\citealt{Vila-Costas:1992,Zaritsky:1994,Kennicutt:2003,Sanchez:2014,Ho:2015,Bresolin:2015}),
but also the myriad of studies concerning the chemical composition of star-forming galaxies, notably those at high redshifts (\eg, to investigate the mass-metallicity relation), that rely on the local calibrations, and the cosmic evolution of metals  (\citealt{Tremonti:2004,Erb:2006,Maiolino:2008,Mannucci:2010,Zahid:2013,Sanders:2015}, to cite only a few).

In order to derive reliable chemical abundances of ionized nebulae it is necessary to have a good knowledge of the physical conditions of the gas, in particular of the electron temperature \te, because of the strong temperature sensitivity of the line emissivities of the various ions. 
An excellent source on the subject of deriving oxygen abundances in ionized nebulae is the monograph published by \citet{Stasinska:2012}.
Nebular electron temperatures can be obtained through the classical, so-called {\em direct}  method\index{direct method} (\citealt{Menzel:1941}), utilizing
emission lines of the same ions that originate from different excitation levels, and in particular from the ratio of the (collisionally excited) auroral \oiii\lin4363 and nebular \oiii\llin4959,\,5007 lines. In the case of high-excitation (low-metallicity) extragalactic \hii\/ regions and planetary nebulae the \oiii\lin4363 line is often detected, but it becomes unobservable as the cooling of the gas becomes efficient at high metallicity, or whenever the objects are faint, so that properly calibrated {\em strong-line} abundance diagnostics\index{strong-line abundance diagnostics} become necessary in order to infer the oxygen abundances. However, even for nebulae where \oiii\lin4363\ can be observed, a poorly understood discrepancy exists between the nebular abundances  based on the direct method  and those obtained from emission line strengths calibrated via photoionization model grids (\citealt{McGaugh:1991, Blanc:2015, Vale-Asari:2016}). A 0.2--0.3 dex discrepancy (\te-based abundances being lower) is also found when using the weak metal {\em recombination} lines, in particular the \oiir\ lines around 4650\,\AA, instead of the collisionally-excited lines (\citealt{Garcia-Rojas:2007a, Esteban:2009, Toribio-San-Cipriano:2016}). On the other hand, comparisons of stellar (B and A supergiants) and nebular chemical compositions in a handful of galaxies (see \citealt{Bresolin:2009a}) provide a generally good agreement when the nebular abundances are calculated from the direct method, at least for sub-Solar metallicities.

Despite the somewhat unsatisfactory situation illustrated above, we can still derive robust results concerning the metallicities of outer disk \hii\ regions. As will become apparent later on, it is important to highlight two results. Firstly, 
radial abundance trends in spiral disks are generally found to be qualitatively invariant relative to the selection of nebular abundance diagnostics, although different methods can yield different gradient slopes (see \citealt{Bresolin:2009a, Arellano-Cordova:2016}). Secondly, 
O/H values that are derived from direct measurements of \te\ or from strong-line diagnostics that are calibrated based on \oiii\lin 4363 detections, lie at the bottom of the possible abundance range, when compared to metallicities derived from other diagnostics, such as those based on theoretical models. 

The literature on nebular abundance diagnostics\index{nebular abundance diagnostics} is vast (a recent discussion can be found in \citealt{Brown:2016}), and for the purposes of this review it is important only to recall a few of the most popular ones, and (some of) their respective calibrations:

\begin{enumerate}

\item\noindent O3N2 $\equiv$ log[(\oiii\lin5007/\hbeta)/(\nii\lin6583/\halpha)],  calibrated empirically (\ie, based on \te-detections in \hii\ regions of nearby galaxies) as given by \citet{Pettini:2004} and, more recently, by \citet{Marino:2013}.\\

\item\noindent N2O2 $\equiv$ \nii\lin 6583/\oii\lin3727, calibrated from photoionization models by \citet{Kewley:2002} and  empirically  by \citet{Bresolin:2007}. \citet{Bresolin:2009} showed that these two  calibrations  yield abundance gradients in spiral disks that have virtually the same slopes, despite a large systematic offset.\\

\item\noindent N2 $\equiv$ \nii\lin6583/\halpha, calibrated by \citet{Pettini:2004} and \citet{Marino:2013}.\\

\item\noindent \rtwothree\ $\equiv$ (\oii\lin3727 + \oiii\llin4959,\,5007)/\hbeta\ (\citealt{Pagel:1979}). Many different calibrations have been proposed through the years (\eg, \citealt{McGaugh:1991, Kobulnicky:2004}).
This diagnostic can be important in order to verify whether the metallicity gradients derived from other strong-line techniques, mostly involving the nitrogen \nii\lin6583 line, are corroborated by considering only oxygen  lines instead. Unfortunately, the use of this indicator for abundance gradient studies is complicated by the non-monotonic behaviour of \rtwothree\ with oxygen abundance. The simultaneous use of a variety of diagnostics, when the relevant emission lines are available, alleviates this problem.
The empirically calibrated P-method (\citealt{Pilyugin:2005a}), and some related diagnostics (\eg, \citealt{Pilyugin:2016}) also make use of  both the \oii\ and \oiii\ strong emission lines.

\end{enumerate}

To summarize, a variety of optical emission-line diagnostics are available to derive the metallicities (oxygen abundances) of extragalactic \hii\ regions. Due to poorly known aspects of nebular physics (\eg, temperature fluctuations), absolute metallicities  remain a matter of debate, especially at high (near Solar) values. On the other hand, relative abundances within galaxy disks are quite robust. \te-based oxygen abundances lie at the bottom of the distribution of values obtained from the set of metallicity diagnostics currently available.

\section{Chemical Abundances of H\,II Regions in Outer Disks}\label{Sect:work}

While the spectroscopic analysis of \hii\ regions in the inner disks of spiral galaxies is a well-developed activity in extragalactic astronomical research, starting with the pioneering work by \citet{Searle:1971} and \citet{Shields:1974}, who provided the first evidence for the presence of exponential radial abundance gradients in nearby galaxies such as M33 and M101, the investigation of ionized nebulae located beyond the boundary of the main star-forming disk has begun only recently.
Observationally, the main difficulty in measuring chemical abundances in these outlying \hii\ regions is represented by their intrinsic faintness. In fact, these nebulae are typically ionized by single hot stars, as deduced from their \halpha\ luminosities (\citealt{Gil-de-Paz:2005, Goddard:2010}), that are on average about two orders of magnitude fainter than those of the giant \hii\ regions that are routinely observed in the inner disks (\citealt{Bresolin:2009,Goddard:2011}). This section reviews the investigations of the chemical abundances in the outskirts of nearby spiral galaxies, focusing mostly on oxygen in \hii\ regions. Studies of metals in old stars and nebular nitrogen are also briefly discussed.

\subsection{Early Work}

Spectra of a handful of outlying \hii\ regions in the disks of the late-type spirals NGC~628, NGC~1058 and NGC~6946, known for their extended \hi\ distributions, were first obtained by \citet{Ferguson:1998a}. These authors found that the oxygen abundance gradients 
measured in the inner disks appear to continue to large galactocentric distances, beyond the isophotal radii\index{isophotal radius}\footnote{At the isophotal radius \rtf\ the surface brightness measures 25\,mag\,arcsec$^{-2}$, and is often reported in the $B$ photometric band, as in the Third Reference Catalogue of Bright Galaxies (\citealt{de-Vaucouleurs:1991}).} \rtf, and reaching low metallicities, corresponding to $10-15$ percent of the Solar value. Unfortunately, the small sample size  concealed the possibility, demonstrated by later work, that the radial metallicity trends could actually be different between inner and outer disks. \citet{van-Zee:1998a} also presented spectra of outlying \hii\ regions in a sample of 13 spirals, but the number of objects observed near \rtf\ and beyond was very small.

The oxygen abundance in the outer disks of two iconic representatives of the class of extended UV (XUV) disk galaxies discovered by the {\it GALEX} satellite, M83 (\citealt{Thilker:2005}) and NGC~4625 ({\citealt{Gil-de-Paz:2005}), was investigated by means of multi-object spectroscopy with the Magellan and Palomar 200-inch telescopes by \citet{Gil-de-Paz:2007}. These authors also found a relatively low metal content, around $10-20$ percent of the Solar value, utilizing a combination of photoionization models and the \rtwothree\ strong-line abundance diagnostic. This study, however, was also limited by the small number of \hii\ regions observed at large galactocentric distances, especially for the galaxy NGC~4625. A feature of the M83 radial abundance gradient that was suggested by \citet{Gil-de-Paz:2007}, \ie, a sudden drop in metallicity at a galactocentric distance of 10\,kpc, 
but whose presence was dubious due to the uncertain O/H ratios derived from \rtwothree,
was also detected later in the  larger sample of outlying \hii\ regions studied by \citet{Bresolin:2009}.
\citet{Gil-de-Paz:2007} remarked that such a sharp decrease, if present, could be the signature of a transition to an outer disk where the star formation efficiency is significantly lower compared to the inner disk.

\subsection{M83: a Case Study}\label{Sec:m83}
The first investigation to obtain robust chemical abundances---via a variety of nebular metallicity diagnostics---in the outer disk of a single galaxy was carried out by \citet{Bresolin:2009}, who obtained  spectra of ionized nebulae in the outer disk of M83\index{M83 (NGC5236)} with the ESO Very Large Telescope. 
Of these \hii\ regions, 32 lie at galactocentric distances larger than the isophotal radius, extending out to 22.3\,kpc (2.64\,\rtf) from the galaxy centre.

The principal chemical abundance properties of the outer disks of spiral galaxies,  confirmed by subsequent investigations of other targets (as discussed in the next pages), are all showcased in this prototypical XUV disk galaxy (see Fig.~\ref{fig:m83}): 

\begin{figure}[t]
\centering
\includegraphics[scale=0.95]{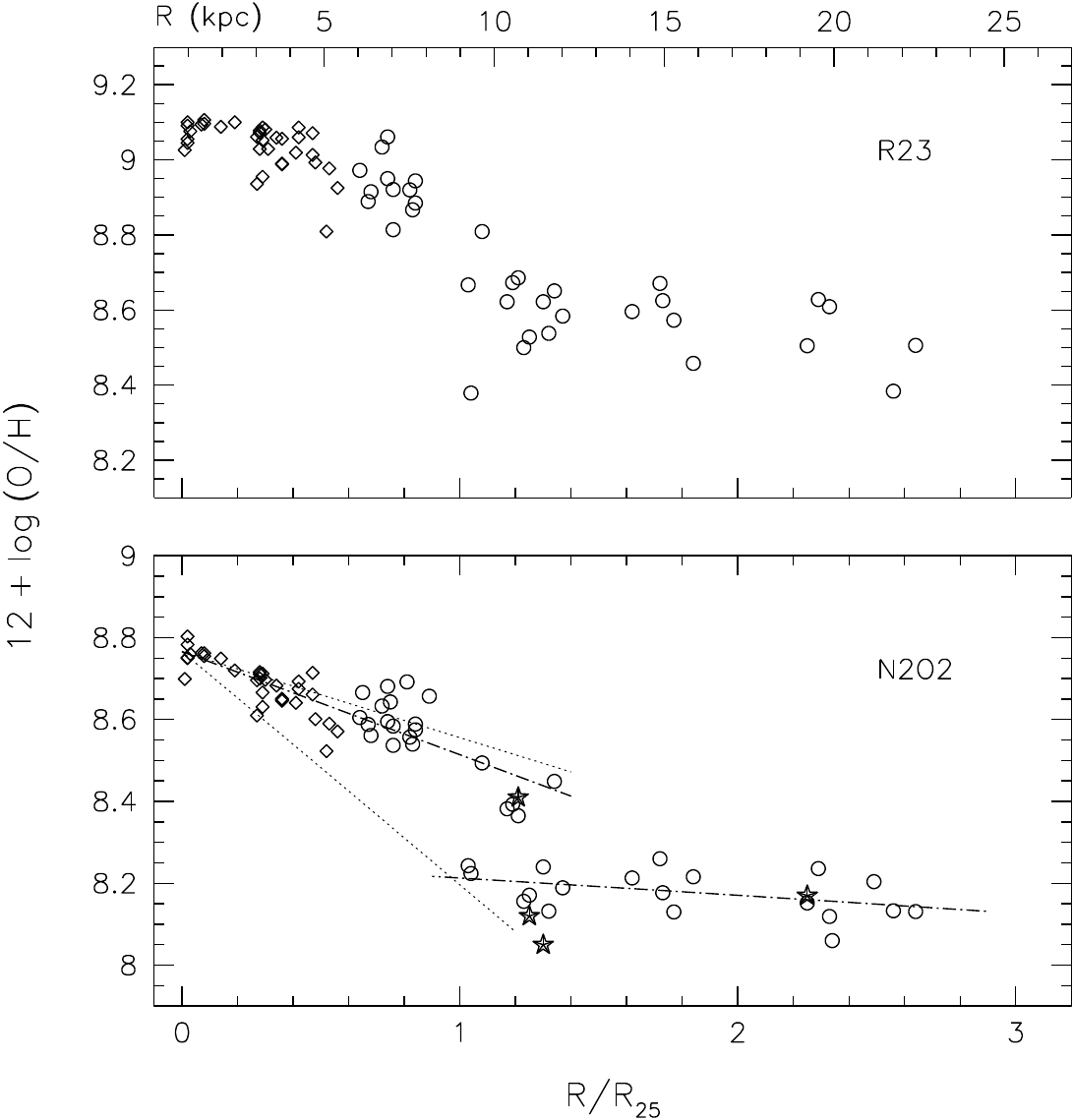}
\caption{The radial oxygen abundance gradient in M83, determined from two different diagnostics (\rtwothree\ in the {\it top} panel, N2O2, as calibrated empirically by \citealt{Bresolin:2007}, in the {\it bottom} panel), from \hii\ regions located in the inner disk ($\diamond$ symbols: \citealt{Bresolin:2002, Bresolin:2005}) and in the outer disk ($\circ$ symbols: \citealt{Bresolin:2009}). The $\star$ symbol represents \oiii\lin4363-based O/H values. Linear regressions to the radial abundance gradient are shown as separate dot-dashed lines for the inner and outer portions of the galactic disks. The two dotted lines represent the range of gradient slopes measured by \citet{Ho:2015} from a sample of 49 galaxies: $\rm d\log(O/H)/dR = -0.39 \pm 0.18$ dex\,\rtf$^{-1}$. Adapted from the data published by \citet{Bresolin:2009}}
\label{fig:m83}       
\end{figure}

\begin{itemize}
\item  the nearly flat radial gradient beyond \rtf, contrasting with the steeper exponential decline observed in the inner disk. The gradient slope found for the inner disk using the N2O2 diagnostic
 
\begin{equation}
{\rm \frac{d\log(O/H)}{dR}} = -0.25 \pm 0.02~ {\rm dex}\,R_{25}^{-1}
\end{equation}

compares well with the benchmark value of $-0.39 \pm 0.18$ dex\,\rtf$^{-1}$ measured by \citet{Ho:2015} from a sample of 49 galaxies (the two dotted lines in Fig.~\ref{fig:m83} show the two extreme values of the one sigma range). On the other hand, in the outer disk a linear fit to the data displayed in Fig.~\ref{fig:m83} yields a slope

\begin{equation}
{\rm \frac{d\log(O/H)}{dR}} = -0.04 \pm 0.02~ {\rm dex}\,R_{25}^{-1},
\end{equation}

\ie, nearly flat. Clearly, the radial behaviour of the gas metallicity differs significantly  between the inner, star-forming disk of M83, and the outer disk.\\

\item  the relatively high mean O/H value measured at large galactocentric distances. This value depends on the selection of abundance diagnostic, as explained in Sect.~\ref{Sec:diagnostics}. In the work presented by \citet{Bresolin:2009} it lies in the range \oh~=~8.2--8.6, \ie, between $\sim$30\% and $\sim$80\% of the Solar value, depending on whether the abundances are tied to \te-based detections (lower value) or whether they are determined from photoionization models (upper value). It is worth pointing out that the few detections of the \oiii\lin4363 auroral line lead to O/H ratios (star symbols in Fig.~\ref{fig:m83}) that are quite consistent with those determined from the empirically calibrated N2O2 diagnostic.

This result on the metallicity of the outer disk of M83 is at odds with the expectation that the very outskirts of spiral galaxies are somewhat pristine and chemically unevolved, as would be implied by a simple picture of inside-out galactic formation. I draw attention to the fact that adopting abundance diagnostics that are calibrated via photoionization models, the outer disk of M83 would have a mean metallicity that is nearly Solar. Using a more conservative approach, we can say that the mean metallicity of the outer disk is {\em at least} 1/3 Solar, based on the summary presented at the end of Sect.~\ref{Sec:diagnostics}.

\citet{Bresolin:2009} also pointed out that the extended disk of M83 should be considered chemically over-enriched given its large gas mass fraction (approaching unity) when compared to a closed box chemical evolution model, the opposite behaviour of what is observed, for example, in dwarf galaxies (\citealt[see also the explanatory text for Eq.~\ref{eq:yeff}, below]{Matteucci:1983}). This point will be further discussed in Sect.~\ref{sec:other}.

\end{itemize}

Fig.~\ref{fig:m83} also suggests the presence of a $\sim$0.2~dex oxygen abundance discontinuity beyond the isophotal radius. This feature is not confirmed by all abundance diagnostics considered (see also \citealt{Pilyugin:2012a}), nor is it detected in other
extended disk galaxies, except NGC~4625 (\citealt{Goddard:2011}), but it also appears in the data presented by \citet{Gil-de-Paz:2007}. It resembles the break occurring for the $\alpha$-elements measured for Cepheids in the Milky Way at a galactocentric distance of approximately 9\,kpc (\citealt{Lepine:2014}).

The observations in M83 demonstrate that spectroscopy of \hii\ regions located in the extended, gas-rich disks of spiral galaxies allows us to probe the present-day chemical abundances of external galactic disks out to nearly three isophotal radii, equivalent---in the case of M83---to more than 20\,kpc. The interesting, and somewhat surprising, result is found that the radial metallicity distribution becomes virtually flat in the extended outskirts, with a value of at least 1/3 of Solar. The extended disk appears to be chemically over-abundant for its very large gas mass fraction.

\subsection{Other Systems}\label{sec:other}
In this Section the results obtained from other investigations of single targets or small samples of galaxies, essentially confirming and expanding the general picture outlined in Sect.~\ref{Sec:m83} for M83, are reviewed. In addition to \hii\ regions as primary probes of the present-day metallicity, information from the older stellar content is included. Some galaxies display a flattening of the gas metallicity already inside the main disk ($R<$\,\rtf), perhaps as a result of gas flows induced by the gravitational potential of a stellar bar (\citealt{Martin:1995, Zahid:2011a, Marino:2012}). Here I will focus on outer disk systems exclusively.

\runinhead{H\,I-selected galaxies} 
The oxygen abundances of outlying \hii\ regions in a sample of 13 \hi-selected galaxies were measured using the \rtwothree\ method by \citet{Werk:2011}. The sample is dominated by interacting systems and galaxies displaying a disturbed, extended neutral gas morphology.  In most cases a flat radial abundance distribution was found across most of the disk of these systems, although the number of \hii\ regions observed per galaxy is sometimes too small to infer variations between the inner and outer disks.
Thus the flattening observed can be of a different nature in this kind of galaxies (see the discussion below) compared with relatively isolated galaxies such as M83, where the flattening is observed to occur only in the outer disk. 

\citet{Werk:2011} showed that the oxygen abundances in the outskirts of the galaxies included in their sample are considerably higher than expected, given their large gas content compared with the total barionic (stars + gas) mass. The same result was described by \citet{Bresolin:2009} and \citet{Werk:2010} for the extended disks of M83 and the blue compact dwarf galaxy NGC~2915, respectively. While for typical star-forming galaxies the effective oxygen yield\index{effective yield}, defined by

\begin{equation}\label{eq:yeff}
\rm {\it y}_{\!\mbox{\scriptsize O,eff}}
= \frac{{\it Z}_O}{\ln(\mu^{-1})},
\end{equation}

where $\mu$ is the gas mass fraction and $Z_{\rm O}$ is the metallicity equivalent to the O mass fraction\footnote{The gas mass fraction is $\mu = M_{\rm gas}/(M_{\rm gas} + M_{\rm stars})$. The O mass fraction (`metallicity') and the abundance by number are linked by the relation $Z_{\rm O} = 11.81$ (O/H). The coefficient of proportionality is calculated as $16\,X$, adopting the hydrogen mass fraction for Solar composition from \citet{Asplund:2009}.}, lies below the theoretical oxygen yield\index{yield} for a stellar population, $y_{\rm O} \simeq 0.007$ (\eg, \citealt{Kobayashi:2006}), in the case of the outer disks the opposite holds, with \yoeff~$>$~0.02 (\citealt{Werk:2011, Lopez-Sanchez:2015}). This in essence implies that the oxygen abundances measured in the gas located in the outskirts of these galaxies, including 
the XUV disk systems, characterized by extended \hi\ envelopes, exceed the  values predicted by the closed-box galactic chemical evolution model. 
How this level of chemical enrichment can be attained, given the low values measured for the star formation rate, will be addressed in Sect.~\ref{Sect:models}.

\runinhead{Interacting systems}

The majority of the galaxies studied by \citet{Werk:2011} are located in interacting systems\index{interacting galaxies}.
Nebular oxygen abundances have been measured in the main star-forming disks and along tidal features of interacting and merging systems by various authors (\eg~\citealt{Rupke:2010a, Rich:2012, Torres-Flores:2014}), finding significantly flatter radial distributions compared to non-interacting systems.
These studies are supported by numerical simulations (\citealt{Rupke:2010, Torrey:2012}), showing that gas flows induced by galaxy interactions redistribute the gas in such a way that the original abundance gradients---present in the galactic disks before the merging process---flatten progressively with merger stage. 

This redistribution and radial mixing of metals can take place over very large distances, in extreme cases reaching several tens of kpc. For example, \citet{Olave-Rojas:2015} measured the chemical abundances of \hii\ regions located along the main tidal tail of NGC~6845A, part of a compact, interacting group of galaxies, out to almost 70\,kpc from the centre (approximately 4\,\rtf). The radial oxygen distribution displays a remarkably shallow gradient.  

An interesting interacting system is represented by NGC~1512, which is experiencing an encounter with the companion galaxy NGC~1510. The system, an XUV disk galaxy, is embedded in a very extended \hi\ envelope, with a radius of 55\,kpc (\citealt{Koribalski:2009}), in which low-level star formation is taking place, as shown by the far-UV and \halpha\ emission originating from low-luminosity stellar complexes and associated \hii\ regions.
The flat and relatively high O/H abundance values, \oh~$\simeq$ 8.3, out to a galactocentric distance of $\sim$30\,kpc, have been studied by \citet{Bresolin:2012} and, more recently, by \citet{Lopez-Sanchez:2015}.
The latter authors point out the effect of the interaction on the outlying northern \hi\ spiral arm, where the O/H values have a much larger dispersion than in the opposite side of the galaxy, which remains relatively undisturbed by the ongoing interaction.

\runinhead{Old stars}
The data discussed so far refer only to the present-day metallicities, as derived from \hii\ region emission. It is also possible to infer the chemical composition of the outer disks of nearby spirals from stellar photometry of older populations, in particular of red giant branch (RGB) stars\index{RGB stars}.
The method requires the photometry of individual stars, and as such has successfully been applied only to nearby systems, out to approximately 3\,Mpc. Care must be taken in interpreting these photometric metallicities when discussing disk radial gradients, because of the potential contamination from halo stars.

The metallicity can be derived from a comparison of the observed stellar colours, such as $V-I$, with theoretical stellar tracks, exploiting the fact that these broad-band colours are more sensitive to metallicity than age. In this way, \citet{Worthey:2005} obtained a flat metallicity gradient in the outer disk of M31, between 20 and 50\,kpc from the galaxy centre (1 to 2.5\,\rtf), with a mean value of $\rm [Z/H] \simeq -0.5$. The published \hii\ region abundances (\eg, \citealt{Zurita:2012, Sanders:2012}) do not extend beyond 25\,kpc, and thus whether a similar behaviour is encountered for the younger stellar populations cannot be verified. A flat gradient, with an approximately Solar O/H value, extending out to $\sim$100\,kpc, has been reported for planetary nebulae\index{planetary nebulae} by \citet{Balick:2013} and \citet{Corradi:2015}. These authors attribute this finding to a star formation burst following interactions and merger processes, perhaps related to an encounter with M33, that occurred approximately 3\,Gyrs ago.

\citet{Vlajic:2009, Vlajic:2011} measured the metallicity of RGB stars from deep $g'$ and $i'$ Gemini photometry in the outer disks of the two Sculptor Group spirals NGC~300 and NGC~7793, out to 15\,kpc (2.3\,\rtf) and 11.5\,kpc (2.4\,\rtf), respectively. Both galaxies, like M31, display purely exponential surface brightness profiles out to these large galactocentric distances, indicating that the halo contribution is probably negligible. The possible connection between gas-phase metallicity and surface brightness profiles will be briefly discussed in Sect.~\ref{Sect:marino}.

For both NGC~300 and NGC~7793 the stellar metallicity flattens out to an approximately constant value, or even slightly increases with radius in the outer disk, in contrast with the exponential decline inferred from \hii\ regions and young stars in the inner disk.
The measured metallicity is quite low, $\rm [Fe/H] \simeq -1$ for the outer disk of NGC~300, and $\rm [Fe/H] \simeq -1.5$ (or even lower, depending on the age of the stars) for NGC~7793, but could be compatible with the present-day metallicity of the inner disk if the chemical enrichment due to stellar evolution between the time probed by the RGB stars ($8-12$\,Gyr ago) and the present epoch is taken into account. 
These results are made somewhat uncertain by the age-metallicity degeneracy, the assumption of a single age for the RGB stars and the potential effects of stellar migration.

\runinhead{Other XUV disks}
The chemical abundances of outer \hii\ regions  in a few XUV disk galaxies\index{XUV disks} (as defined in the catalogue by \citealt{Thilker:2007}), in addition to M83 and NGC~1512, have been presented by different authors. 
For convenience, Table~\ref{tab:1} summarizes these studies.
These investigations differ somewhat in spectroscopic depth, abundance diagnostics adopted and radial coverage, but they tend to provide a unified picture regarding the abundance gradients, in particular the presence of a break occurring approximately at the isophotal radius, as a dividing point between the inner disk, characterized by an exponential nebular abundance gradient, and the outer disk, with a shallower or flat abundance gradient.

\begin{table}
\caption{Nebular abundance studies in XUV disks}
\label{tab:1}       
\begin{tabular}{p{3.5cm}p{4cm}p{3cm}}
\hline\noalign{\smallskip}
Galaxy	& References	& Largest radius  \\
		&				& (\rtf\ units)  \\
\noalign{\smallskip}\svhline\noalign{\smallskip}
NGC 628			& Rosales-Ortega et al.~2011$^a$	& 1.7 \\ \nocite{Rosales-Ortega:2011}
NGC 1512 		& Bresolin et al.~2012				& 2.2 \\
		 		& L\'opez-S\'anchez et al.~2015		& 2.8 \\
NGC 3621 		& Bresolin et al.~2012				& 2.0 \\
NGC 4625 		& Goddard et al.~2011				& 2.8 \\ 
NGC 5253 (M83)	& Bresolin et al.~2009				& 2.6 \\
\noalign{\smallskip}\hline\noalign{\smallskip}
\end{tabular}

$^a$ Data for  objects lying beyond \rtf\ extracted from \citet{Ferguson:1998a}.
\end{table}

Not included in Table~\ref{tab:1} is NGC~3031 (M81), for which a flat outer gradient has been suggested, but this result relies on a very small sample of outlying \hii\ regions (\citealt{Patterson:2012, Stanghellini:2014}). A more recent work by \citet{Arellano-Cordova:2016} does not find evidence for a flat  gradient out to a galactocentric distance of 33\,kpc (2.3\,\rtf}). This seems to be consistent with the  shallow overall abundance gradient, both in dex\,kpc$^{-1}$ and normalized to the isophotal radius, which is possibly the consequence of galaxy interactions. 
It is also worth pointing out that the abundance break observed in NGC~3621 by \citet{Bresolin:2012} has been confirmed by the independent spectroscopic analysis of five blue supergiant stars, straddling the isophotal radius, by \citet{Kudritzki:2014}. The stellar metallicities are intermediate between the nebular metallicities determined from the N2 and R23 diagnostics. Finally, it is important to notice that the sample presented above includes fairly isolated systems (NGC~3621, M83, NGC~628), ruling out the possibility that  abundance breaks and significant metal mixing develop only as a consequence of recent galaxy interactions.

\runinhead{The Milky Way}
Evidence for a flattening of the abundance gradient in the outer disk of the Milky Way comes from  observations of various metal tracers, which also sample populations with different ages: Cepheid variables (\citealt{Korotin:2014}), open clusters ({\citealt{Magrini:2009b,Yong:2012}), and \hii\ regions (\citealt{Vilchez:1996,Esteban:2013}). This break appears at a galactocentric distance around 12\,kpc, extending outwards to 19--21\,kpc, as shown from either Cepheids (\citealt{Genovali:2015}) or open clusters (\citealt{Carraro:2004}). While the flattening in the Cepheids chemical abundances is still somewhat controversial (\eg, \citealt{Lemasle:2013}), the open clusters show a clear bimodal radial gradient in metallicity ({\citealt{Yong:2012}), the outer gradient being quite shallow, with a characteristic outer disk metallicity [Fe/H]~$\simeq$~$-0.3 \pm 0.1$. Further studies of the behaviour of the radial distribution of the stellar metallicity (and chemical element patterns) in the outer disk of the Galaxy will be important to constrain models of the chemical evolution of the Milky Way, and the effects of the corotation resonance and stellar radial migration (\citealt{Mishurov:2002,Lepine:2011, Korotin:2014}).

\subsection{Results from Galaxy Surveys}

More recent results about the chemical abundances of the outer disks of spiral galaxies in the nearby Universe have been published as part of relatively large spectroscopic surveys, largely dedicated to the measurement of emission-line abundances of the interstellar medium in the parent galaxies. These surveys, based on 4\,m-class telescope observations, do not reach emission line levels
as faint as those probed by some of the single galaxy work illustrated earlier. Therefore, weak lines such as \oiii\lin 4363 remain undetected in the low-luminosity \hii\ regions located in the galactic outskirts. In addition, these surveys have provided metallicity information 
out to $\sim$1--1.5\,\rtf, \ie, to considerably smaller galactocentric distances than possible with 8\,m-class facilities (see Table~\ref{tab:1}).
On the other hand, the large number of galaxies (hundreds) provides essential statistical information about the properties of the abundance gradients, that are necessary to establish, for instance, how common radial metal distribution breaks are within the general population of spiral galaxies. Furthermore, such surveys also enable the investigation of  possible 
correlations between abundance gradients and galactic attributes, such as mass, star formation rate, and structural properties (\eg, the presence or absence of bars).

\runinhead{Integral Field Spectroscopy}
\citet{Sanchez-Menguiano:2016} presented oxygen abundance measurements obtained by the Calar Alto Legacy Integral Field Area (CALIFA) project (\citealt{Sanchez:2012})
in 122 face-on spiral galaxies. Adopting the O3N2 nebular diagnostic, they confirmed earlier results, obtained from the same survey (\citealt{Sanchez:2014}), that a flattening of the gas-phase oxygen abundance taking place around a galactocentric distance corresponding to twice their effective radii\index{effective radius}\footnote{The effective radius \re\ encloses 50\% of the light, integrated by adopting an exponential radial profile of the surface brightness (\ie, not including the contribution from the bulge): $I=I_0 \exp[-(R/R_{\mathrm d})]$, with $I_0$ the central intensity and $R_{\mathrm d}$ the disk scale-length. The effective radius is given by $R_{\mathrm e} = 1.678\,R_{\mathrm d}$ (\eg, \citealt{Graham:2005}).}
(\re, measured in the $r$ band) is a common occurrence in spiral disks. About 82\% of the sample with reliable abundance data in the outer disks show this effect, with no apparent dependence on galactic mass, luminosity, and morphological type. The oxygen abundance in the inner disks, on the other hand, follows a  gradient having a characteristic slope of approximately $-0.07$\,dex\,\re$^{-1}$, except for the very central parts (\citealt{Sanchez:2014}). This common behaviour is illustrated in Fig.~\ref{fig:califa}, which displays data extracted from \citet[their Fig.~9]{Sanchez-Menguiano:2016}.

In order to make these results more easily comparable with those presented earlier, where the radial normalization is done relative to the isophotal radius, we need to define a relation between \re\ and \rtf, which depends on the central surface brightness value ($\mu_0$) for the adopted exponential brightness profile. Taking $\mu_0 = 21.65$\,mag\,arcsec$^{-2}$ from \citet{Freeman:1970}, and using $\mu(R) = \mu_0 + 1.086 R/R_d$, one obtains 
\rtf\,=\,1.84\,\re. Thus, the flattening in the abundance gradient observed for the CALIFA sample of galaxies to occur at $R \sim 2$\,\re\ or, equivalently, around $R \sim $\,\rtf, is consistent with what is reported in Sect.~\ref{Sec:m83} and \ref{sec:other}. The O3N2-based oxygen abundances measured in the outer disks, out to $\sim$~1.5\,\rtf, are also roughly consistent with those presented earlier, in particular they represent a significant fraction of the Solar value, \eg, approximately 0.5\,(O/H)$_\odot$ for the intermediate mass bin shown in Fig.~\ref{fig:califa}.

\begin{figure}
\includegraphics[scale=0.95]{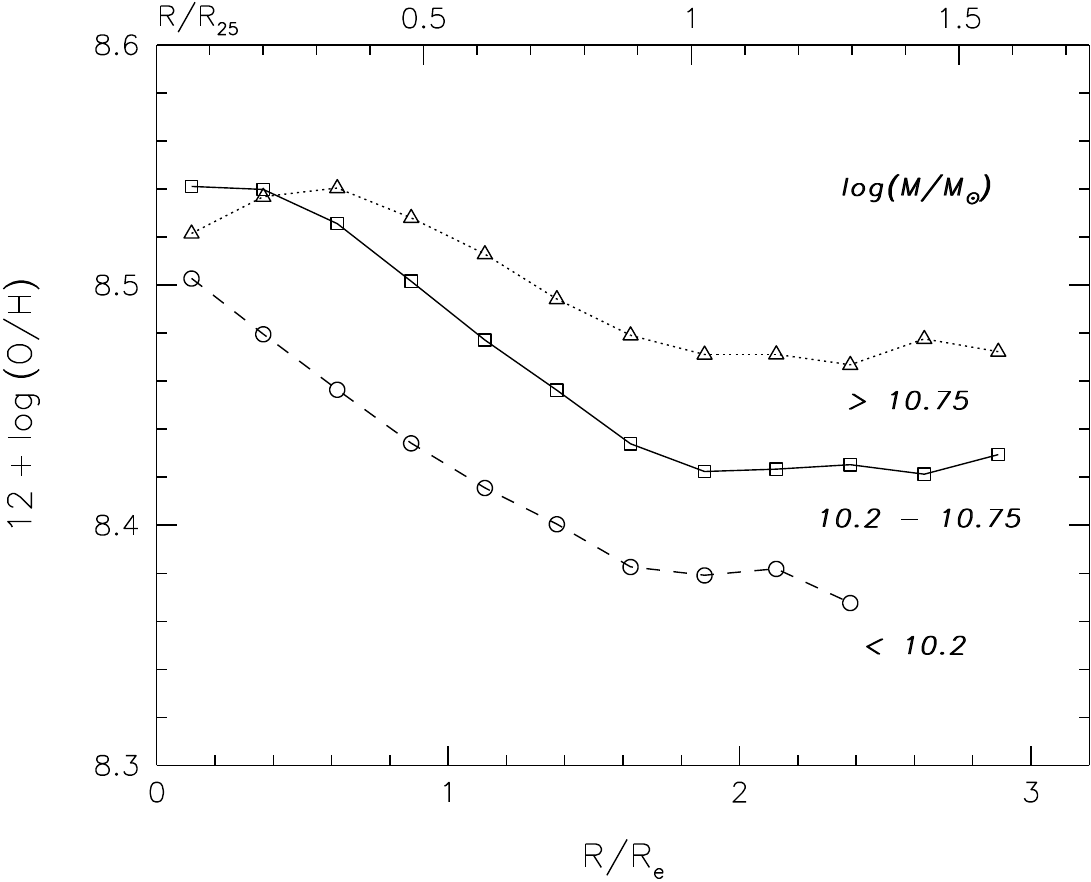}
\caption{Mean radial oxygen abundance profiles measured for a sample of face-on spirals from the CALIFA survey. The data are plotted in bins of 0.25\,\re, for three different galaxy mass ranges, as indicated in the plot.
The upper scale, drawn in units of the isophotal radius, assumes a central surface brightness $\mu_0 = 21.65$\,mag\,arcsec$^{-2}$. Adapted from \citet[Fig.~9]{Sanchez-Menguiano:2016}}
\label{fig:califa}       
\end{figure}

\runinhead{Long-slit spectra}
For completeness, it is worth mentioning some additional surveys that obtained spectroscopic observations of the ionized gas in regions close to the edges of spiral galaxies, even though the radial coverage is not as extended as in the cases discussed so far, and is generally limited to regions inside the isophotal radius.
\citet{Moran:2012} obtained long-slit spectra along the major axis of 174 star-forming galaxies from the \{it GALEX} Arecibo Sloan Digital Sky Survey (\citealt{Catinella:2010})
with stellar mass $M > 10^{10}\,M_\odot$, and determined O3N2-based gas-phase oxygen abundances in spatial bins for 151 galaxies displaying emission lines. However, their data extend to galactocentric distances of about 1.5\,$R_{90}$\footnote{$R_{90}$ encloses 90\% of the galaxy light, including the bulge.}, or approximately 0.9\,\rtf\ according to the transformation between the two normalization radii estimated by these authors. Thus, these chemical abundances still refer to the main star-forming disk, and should not be compared directly with the outer disk abundance properties presented in the previous sections. Interestingly, however, for about 10\% of their galaxies \citet{Moran:2012} measured a significant drop in O/H around $R=R_{90}$, whose magnitude correlates with the total \hi\ mass fraction. These authors suggest that the downturn in oxygen abundance results from the accretion of relatively metal-poor gas in the outer regions of these galaxies.

Similar long-slit observations have been carried out by \citet{Carton:2015} for 50 \hi-rich galaxies, 
part of the Bluedisk survey (\citealt{Wang:2013}),
with a radial coverage extending to about 2\,$R_{90}$ in some cases.
Also for these targets a steepening of the radial abundance distribution is observed at large radii. However, in this work the oxygen abundance downturn is not found to correlate with the \hi\ properties of the parent galaxies as instead found in the work by \citet{Moran:2012}.

\subsection{Nitrogen Abundances}
The investigation of nitrogen abundances\index{nitrogen abundance} in extragalactic \hii\ regions, and in particular of the N/O abundance ratio, provides important constraints on the chemical evolution of galaxies. This springs from the fact that, while oxygen is the nucleosynthetic product of massive stars ($M > 8 M_\odot$), nitrogen, whose abundance can  in general be easily measured in nebular spectra\footnote{The N/O abundance ratio is obtained from the \nii\llin6548,6583/\oii\lin3727 line ratio, using the commonly adopted ionization correction scheme N$^+$/O$^+$~=~N/O.},  originates mostly in intermediate-mass ($M = 1-8 M_\odot$) stars (\citealt{Henry:1999}). 
A look at the N/O ratio variation as a function of metallicity O/H in extragalactic nebulae reveals a bimodal behaviour. The N/O ratio is approximately constant [log(N/O)~$\simeq -1.4$, but with a large scatter, see \citealt{Garnett:1990}] below \oh~=~8.0, and increases with O/H at larger metallicities. This is interpreted in terms of a primary production of nitrogen, which is what is predominantly being measured at low metallicity, and of a secondary component, proportional to the oxygen abundance, dominating at high O/H  (\citealt{Vila-Costas:1993}). The N/O ratio measured in outer disk \hii\ regions conforms to this trend, as shown by \citet{Bresolin:2012}. Since the oxygen abundances measured in outer spiral disks are generally below the level at which secondary nitrogen production becomes predominant, the radial trend of the N/O abundance ratio is virtually flat, with log(N/O)~$\simeq -1.3$ to $-1.5$, in these outer regions (\citealt{Bresolin:2009, Berg:2012, Lopez-Sanchez:2015}). 
A similar behaviour has recently been observed by \citet{Croxall:2016} in M101 (also an XUV galaxy, \citealt{Thilker:2007}), with the onset of the flattened N/O radial distribution occurring around 0.7\,\rtf. These results stress the fact that primary production of nitrogen dominates in the \hii\ regions populating the outskirts of spiral galaxies.

In summary, flat abundance gradients and relatively high oxygen abundances appear to be common features of star-forming outer disks.

\section{Additional Considerations}\label{Sect:add}
To conclude the discussion of the observational constraints on the metallicity of the outer regions of spiral galaxies I include two additional topics that have been addressed by recent work. Future investigations of the chemical abundance properties of the outer disks of spiral galaxies will benefit from the study of the spatially-resolved gas content (both atomic and molecular), which will help to shed light on the interplay between chemical and secular evolution of the outer disks and accretion events or gas flows taking place in the very outskirts of galaxies.

\subsection{Relation Between Metallicity and Surface Brightness Breaks}\label{Sect:marino}
Recent work by \citet{Marino:2016} probed into the possible connection between outer disk abundance gradients and  surface brightness profile breaks that characterize the disks of spiral galaxies, as discussed elsewhere in this volume. These authors focused on 131  galaxies, extracted from the larger CALIFA sample, displaying either Type~II (`down-bending') or Type~III (`up-bending') surface brightness profiles. A correlation was found in the case of Type~III galaxies. At lower masses, $\log(M/M_\odot) < 10$, a modest flattening in the $g'-r'$ colour tends to be a common feature, together with  a mild flattening of the O/H gradient, while at higher masses both colour and O/H radial profiles display a pronounced flattening. The different behaviour detected for the Type~III galaxies is tentatively attributed by \citet{Marino:2016} to a downsizing effect, such that the higher-mass systems have already experienced a phase of inside-out growth, while for the smaller systems an enhanced disk buildup phase is more recent or still on-going.

\subsection{An Analogy with Low Surface Brightness Galaxies?}\label{sec:lsbg}
It has been pointed out by some authors (\citealt{Thilker:2007, Bresolin:2009}) that the structural parameters and the star-formation properties of outer spiral disks, such as the low mass surface densities and low star formation rates, resemble those observed in low surface brightness (LSB) galaxies\index{low surface brightness galaxies}. We can then ask the question whether this analogy extends to the chemical abundance properties. In particular, 
does a low star formation efficiency (\citealt{Wyder:2009}) lead to a flat abundance distribution also in the case of LSB galaxies? The question remained without a clear answer until recently, because very few studies addressed the gas-phase chemical abundance properties of this type of galaxies, and in particular their abundance gradients, in part due to observational challenges.  \citet{Bresolin:2015} measured \hii\ region oxygen abundances for a sample of 10 LSB spiral galaxies, and investigated the presence of radial abundance gradients in this sample. They found that LSB galaxies do display radial abundance gradients that, when normalized by the effective radii, are consistent with those measured for high surface brightness galaxies. Thus, the analogy between LSB galaxies and the outer disks of spiral galaxies does not seem to extend to the chemical abundance properties, despite the similarities outlined above. This result suggests that the chemical evolution of LSB galaxies proceeds in a similar fashion to the high surface brightness galaxies, albeit  at a slower pace due to the lower star formation rates, while the outer disks probably follow a different evolutionary path. The latter possibility is addressed in the following section.

\section{The Evolutionary Status of Outer Disks}\label{Sect:models}
A variety of mechanisms can be invoked in order to explain the data presented in the previous section within a coherent picture of galactic chemical evolution. Unfortunately, the theoretical framework is, to a great extent, still lacking, since detailed modelling  accounting for the gas-phase chemical abundances measured in the outer disks has not been developed yet. The relatively high metal enrichment observed in these \hi-rich, low-star formation rate (SFR) regions of spiral galaxies suggests that some form of mixing mechanism, and perhaps more than one, should be responsible for the observed chemical abundance properties of the outer disks.
This section presents some of the possible processes discussed in the literature that could redistribute metals produced in the main star-forming disk of spirals into their very outskirts, tens of kpc from the galactic centres.
The effects of galaxy interactions and merging on the radial abundance gradients have already been introduced in Sect.~\ref{sec:other}, so that the focus will now be on mixing mechanisms that could affect, in principle, also isolated systems, but we should keep in mind that the chemical abundances in the outer disks could be sensitive to encounters that might have occurred in the distant past, as evidenced, for example, by the presence of warps in the \hi\ envelopes. 

The stellar radial migration process (\citealt{Sellwood:2002, Roskar:2008b}) does not affect the present-day abundance gradient of oxygen, since this element, whose abundance we trace with \hii\ region spectroscopy, is produced by massive stars, that do not have sufficient time to migrate radially before ending their lives (\citealt{Kubryk:2015}).
Older tracers of ionized gas metallicity, such as planetary nebulae, can have a different behaviour (\citealt{Magrini:2016}).

\subsection{Flattening the Gradients}
While some of the processes discussed below can explain the  flattening of the \hii\ region oxygen abundances observed in the outer disks, we start with the remark made by Bresolin et al.~(2012), who suggested that the flat O/H distribution could simply be a consequence of relatively flat star formation efficiencies (\citealt{Bigiel:2010,Espada:2011}). Defining the star formation efficiency\index{star formation efficiency} as SFE = $\rm \Sigma_{SFR}/\Sigma_{HI}$ (\eg, \citealt{Bigiel:2008}), \ie, the ratio between the surface densities of  star formation rate (inferred for example from far-UV observations) and \hi\ mass, we can approximate the gas-phase oxygen abundance per unit surface area of the disk, neglecting effects such as gas flows and variable star formation rates, as

\begin{equation}\label{eq:sfe}
\rm \frac{O}{H} \sim \frac{{\it y}_{O}}{11.81}~ {\it t}~ \frac{\Sigma_{SFR}}{\Sigma_{HI}} \propto SFE,
\end{equation}

where $t$ is the duration of the star formation activity, since the amount of oxygen produced per unit surface area and unit time is the product of the oxygen yield\index{yield} (by mass) $y_{\rm O}$ and the star formation rate surface density.
Then, according to (\ref{eq:sfe}) the flattened SFE radial profiles traced beyond the isophotal radius, relative to the behaviour in the inner disks, would result in a similarly flattened O/H radial gradient at large galactocentric radii, as observed.
This appears to be consistent with the idea that the star formation activity at large galactocentric distances proceeds slowly enough  for some metal mixing processes (discussed below) to efficiently  erase or reduce chemical abundances inhomogeneities and large-scale gradients.

Equation~(\ref{eq:sfe}) can also be used to estimate that, given the low star formation rates measured in the outer disks of spiral galaxies and the large \hi\ content, the timescale necessary to reach the observed metal enrichment can be longer than the star formation timescale within an inside-out scenario for galaxy growth or even longer
than a Hubble time, reinforcing the notion that the metallicities measured in outer disks exceed the values attainable by in situ star formation alone (see also Eq.~3 in \citealt{Kudritzki:2014} for an alternative calculation based on the closed box model, and leading to the same conclusion).

The link between SFE and O/H described above has been shown by recent tailored chemical evolution models to be able to reproduce the flattened gas-phase abundances in the outer disk of the Milky Way beyond 10\,kpc from the centre (\citealt{Esteban:2013}) and the flat oxygen abundance gradient in the outer disk of M83 (\citealt{bresolin:2016}, with an adaptation of the chemical evolution model by \citealt{Kudritzki:2015}). This seems also to be consistent with chemical evolution models of the Milky Way in which the decreasing star formation efficiency with increasing galactocentric distance  leads to a flattening of the metallicity gradients in the outer regions (\citealt{Kubryk:2015}).

\subsection{Bringing Metals to the Outer Disks}\label{Sect:mix}
The transport of metals produced in the inner disks to large galactocentric radii appears to be necessary in order to explain the  relatively high gas-phase metallicities observed in the extended disks of spiral galaxies. A number of different mechanisms have been discussed in the literature.
The argumentation contained in the following section is rather speculative since, as already mentioned, a solid theoretical explanation for the chemical properties of extended disks is currently still missing. 
The mechanisms invoked can be broadly divided into two main categories:  mixing and enriched infall. These are succinctly presented below, following in part \citet{Bresolin:2009}, \citet{Bresolin:2012} and \citet{Werk:2011}, to which the reader is referred to for a more in-depth discussion.

\subsubsection{Mixing}
Under this category we can include processes that can be effective in redistributing metals across galactic disks\index{mixing}. Some examples are listed below.

\begin{itemize}

\item Outward radial flows originating from {\em viscosity}\index{viscosity} in the gas layer due, for example, to cloud collisions or gravitational instabilities (\citealt{Lacey:1985, Clarke:1989}) can produce flat gas-phase abundances in the outer disks (\citealt{Tsujimoto:1995, Thon:1998}).\\

\item Gas flows can also be driven by angular momentum redistribution from {\em non-axi\-symmetric structures}, \ie, bars and spiral arms. While stellar bars can affect the chemical evolution in the inner regions of galaxies (\citealt{Athanassoula:1992, Cavichia:2014}), the overlap of spiral and bar resonances\index{resonances} (\citealt{Minchev:2011}) can affect the distribution and metallicity of stars at large galactocentric distances. The resonance associated with the corotation of the spiral pattern has been found to correlate with the radial position of 
breaks in the metallicity gradient by \citet{Scarano:2013}. This suggests that the gas flows occurring in opposite directions, inwards inside corotation and outwards beyond corotation, are connected to the different abundance gradients of the outer disks relative to the inner disks.\\

\item Interstellar {\em turbulence}\index{turbulence} plays an important role in homogenizing the metallicity distribution in galactic disks (\citealt{Scalo:2004}). What constitutes the main driving source, either stellar feedback or gravitational instability, is still uncertain (\citealt{Krumholz:2016}), although in the outer disks, which are characterized by very low star formation rates, it is unlikely that feedback from supernova explosions represents the main source of gas turbulence. The simulations by \citet{Yang:2012} indicate that turbulence driven by thermal instability is quite efficient in erasing kpc-scale metallicity gradients, on timescales that are on the order of a few orbital periods (a few 100~Myrs, \citealt{Petit:2015}). Metals can be transported via convective motions of the gas over large distances, which leads to an equilibrium  between star formation and turbulent mixing,  
making this an appealing mechanism for the explanation of the abundance properties of the extended disks of spirals.

\end{itemize}

\subsubsection{Enriched Infall}\index{enriched infall}
The star formation and chemical enrichment histories of galaxies are profoundly affected by inflows and outflows of gas. Feedback-driven galactic winds\index{galactic winds} eject a large portion of the metals produced in disk stars into the halos, the circumgalactic medium and the intergalactic medium (\citealt{Kobayashi:2007, Lilly:2013, Cote:2015}) out to distances of the order of 100\,kpc (\citealt{Tumlinson:2011, Werk:2013}). These outflows are crucial to explain, for example, the existence of the mass-metallicity relation observed for star-forming galaxies (\citealt{Tremonti:2004, Finlator:2008}).

The gas that has been metal-enriched by supernova explosions at early epochs and subsequently ejected from galaxies can later be re-accreted in a wind-recycling process (see, \eg, the models by \citealt{Oppenheimer:2008, Dave:2011a}). This re-accretion on the disk from the halo should take place preferentially in the outskirts, leading to an inside-out growth, on timescales on the order of a few dynamical times, $\sim$\,1\,Gyr (\citealt{Fu:2013}), necessary for the gas to cool down from the hot phase.
Some evidence in support of this process has been presented recently by \citet{Belfiore:2016} from the spatially resolved metal budget in NGC~628.

\citet{Fu:2013} estimated the gas-phase metallicity of the infalling gas to be around 0.4$\times$~Solar for a Milky Way-type galaxy, which is in rough agreement with the observed metallicity of the extended disks.
It is also worth pointing out that the metallicity of the circumgalactic medium at $z<1$, as traced by Lyman limit systems, is bimodal, with a metal-rich branch peaking at a metallicity approximately 0.5$\times$~Solar, as shown by \citet{Lehner:2013}. According to these authors this metal-rich branch could be tracing cool, enriched gas originating from galactic outflows and tidally stripped material.

The effects of an enriched infall process on galactic chemical evolution models have already been illustrated before. For example, \citet{Tosi:1988} showed how a metal-rich infall would affect the chemical composition of the outer parts of spirals, inducing a flattening of their abundance gradients, and estimated an upper limit for the infalling gas metallicity of 0.4$\times$~Solar from comparisons with the chemical abundances observed in the Milky Way.
Fig.~\ref{fig:m83-model} shows a model radial oxygen abundance gradient for the disk of M83, calculated 
with a galactic wind launched in the inner disk, following \citet{Kudritzki:2015}, but allowing for an inflow of metal-enriched gas with an oxygen abundance \oh\,=\,8.20 (equivalent to 0.32$\times$~Solar). Such an enriched infall is required by the model to reproduce the gas metallicity observed in the extended disk of M83, with the flat distribution arising from the assumed constant star formation efficiency.

\begin{figure}
\includegraphics[scale=0.95]{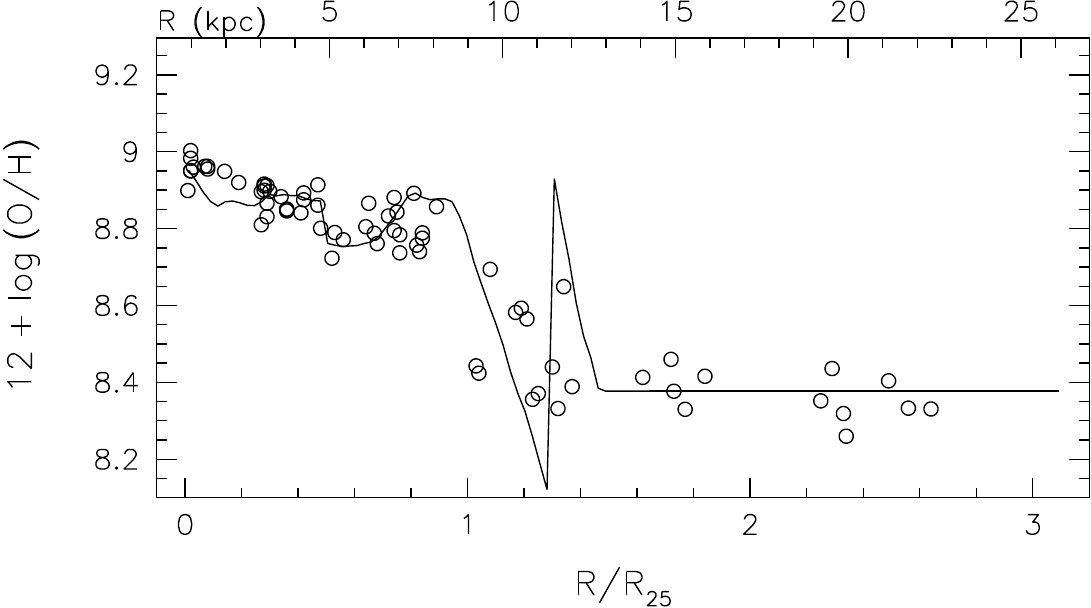}
\caption{Model radial oxygen abundance gradient for the disk of M83 ({\it continuous line}), calculated including an enriched gas infall, compared with the \hii\ region metallicities ({\it open circles}) shown in Fig.~\ref{fig:m83} (\citealt{bresolin:2016})  }
\label{fig:m83-model}       
\end{figure}

\runinhead{Minor mergers}
Minor merger\index{minor mergers} activity, as a source of cold gas leading to mass growth in galaxies, has also been proposed to be effective at chemically enriching the outer disks (\citealt{Lopez-Sanchez:2015}), and is included in this section because its effects could resemble those described above for enriched infall. Given the low accretion rates of star-forming galaxies due to mergers with low-mass satellites measured in the local Universe (\citealt{Sancisi:2008, Di-Teodoro:2014}), this process is unlikely to be important for the chemical evolution of outer disks at the present time. However, higher merger rates in the past (around a redshift $z \simeq 2$) could have made this process a potential contributor to the accretion of metals in the outskirts of galaxies  earlier on during their evolution (\citealt{Lehnert:2016}). Numerical simulations by \citet{Zinchenko:2015} also indicate that the stellar migration process induced by minor merging in Milky Way-type spirals cannot generate the  flattening of the metallicity observed in the outer disks.

Different mechanisms can be invoked to explain the chemical abundance properties of outer disks. Among these are various mixing processes, including turbulence, and metal-enriched infall of gas from the circumstellar medium.

\section{Conclusion}
The outer disks of spiral galaxies remain a relatively unexplored territory in studies of the evolution of galaxies. This Chapter has highlighted the somewhat unexpected attributes of the ionized gas chemical composition discovered in recent years in the outermost parts of galaxy disks, at least for systems with extended \hi\ envelopes and ongoing star formation. At the same time, it is important to stress that metallicity information in the outer disks, gathered until now almost exclusively from \hii\ regions, can provide crucial constraints for models of the chemical evolution of galaxies, considering that these are the most recently assembled regions of the disks according to the inside-out scenario. Future work will more firmly establish how common shallow or flat outer gradients with relatively high oxygen abundances are in spiral disks, which is relevant to ascertain the roles played by enriched galactic inflows and 
mixing mechanisms, such as turbulence, in regulating the chemical evolution of galaxies.
Studies of the evolution of galaxies will benefit from probing the relationship between the gas phase chemical abundances and the properties of the faint stellar populations present in these low surface brightness structures. A better understanding of the mechanisms leading to the formation of these extended structures, and of the importance of the galactic environment in this context is highly desidered. From the observational point of view, the combination of deep integral field spectroscopy with spatially resolved radio \hi\ mapping for large samples of spirals will yield a better characterization of the relationship between gas, star formation, environment and metal production in determining the evolutionary status of present-day galaxies out to very large radii.

\begin{acknowledgement}
The author is grateful to Rob Kennicutt, Emma Ryan-Weber and Rolf Kudritzki for interesting collaborations and stimulating discussions over the years, and to the editors of this volume for the invitation to contribute this Chapter.
\end{acknowledgement}

\bibliographystyle{x_fb}

\bibliography{fb}

\begin{thebibliography}{140}
\providecommand{\natexlab}[1]{#1}

\bibitem[{{Arellano-C{\'o}rdova} et~al.(2016){Arellano-C{\'o}rdova},
  {Rodr{\'{\i}}guez}, {Mayya} and {Rosa-Gonz{\'a}lez}}]{Arellano-Cordova:2016}
{Arellano-C{\'o}rdova}, K.~Z., {Rodr{\'{\i}}guez}, M., {Mayya}, Y.~D.,
  {Rosa-Gonz{\'a}lez}, D.: {The oxygen abundance gradient in M81 and the
  robustness of abundance determinations in H II regions}. \mnras \textbf{455},
   2627--2643 (2016)

\bibitem[{{Asplund} et~al.(2009){Asplund}, {Grevesse}, {Sauval} and
  {Scott}}]{Asplund:2009}
{Asplund}, M., {Grevesse}, N., {Sauval}, A.~J., {Scott}, P.: {The Chemical
  Composition of the Sun}. \araa \textbf{47},  481--522 (2009)

\bibitem[{{Athanassoula}(1992)}]{Athanassoula:1992}
{Athanassoula}, E.: {The existence and shapes of dust lanes in galactic bars}.
  \mnras \textbf{259},  345--364 (1992)

\bibitem[{{Balick} et~al.(2013){Balick}, {Kwitter}, {Corradi} and
  {Henry}}]{Balick:2013}
{Balick}, B., {Kwitter}, K.~B., {Corradi}, R.~L.~M., {Henry}, R.~B.~C.:
  {Metal-rich Planetary Nebulae in the Outer Reaches of M31}. \apj
  \textbf{774},  3 (2013)

\bibitem[{{Belfiore}, {Maiolino} and {Bothwell}(2016)}]{Belfiore:2016}
{Belfiore}, F., {Maiolino}, R., {Bothwell}, M.: {Galaxy gas flows inferred from
  a detailed, spatially resolved metal budget}. \mnras \textbf{455},
  1218--1236 (2016)

\bibitem[{{Berg} et~al.(2012){Berg}, {Skillman}, {Marble}, {van Zee},
  {Engelbracht}, {Lee}, {Kennicutt}, {Calzetti}, {Dale} and
  {Johnson}}]{Berg:2012}
{Berg}, D.~A., {Skillman}, E.~D., {Marble}, A.~R., {van Zee}, L.,
  {Engelbracht}, C.~W., {Lee}, J.~C., {Kennicutt}, R.~C., Jr., {Calzetti}, D.,
  {Dale}, D.~A., {Johnson}, B.~D.: {Direct Oxygen Abundances for Low-luminosity
  LVL Galaxies}. \apj \textbf{754},  98 (2012)

\bibitem[{{Bigiel} et~al.(2010){Bigiel}, {Leroy}, {Walter}, {Blitz}, {Brinks},
  {de Blok} and {Madore}}]{Bigiel:2010}
{Bigiel}, F., {Leroy}, A., {Walter}, F., {Blitz}, L., {Brinks}, E., {de Blok},
  W.~J.~G., {Madore}, B.: {Extremely Inefficient Star Formation in the Outer
  Disks of Nearby Galaxies}. \aj \textbf{140},  1194--1213 (2010)

\bibitem[{{Bigiel} et~al.(2008){Bigiel}, {Leroy}, {Walter}, {Brinks}, {de
  Blok}, {Madore} and {Thornley}}]{Bigiel:2008}
{Bigiel}, F., {Leroy}, A., {Walter}, F., {Brinks}, E., {de Blok}, W.~J.~G.,
  {Madore}, B., {Thornley}, M.~D.: {The Star Formation Law in Nearby Galaxies
  on Sub-Kpc Scales}. \aj \textbf{136},  2846--2871 (2008)

\bibitem[{{Blanc} et~al.(2015){Blanc}, {Kewley}, {Vogt} and
  {Dopita}}]{Blanc:2015}
{Blanc}, G.~A., {Kewley}, L., {Vogt}, F.~P.~A., {Dopita}, M.~A.: {IZI:
  Inferring the Gas Phase Metallicity (Z) and Ionization Parameter (q) of
  Ionized Nebulae Using Bayesian Statistics}. \apj \textbf{798},  99 (2015)

\bibitem[{{Bresolin}(2007)}]{Bresolin:2007}
{Bresolin}, F.: {The Oxygen Abundance in the Inner H II Regions of M101:
  Implications for the Calibration of Strong-Line Metallicity Indicators}. \apj
  \textbf{656},  186--197 (2007)

\bibitem[{{Bresolin}, {Garnett} and {Kennicutt}(2004)}]{Bresolin:2004}
{Bresolin}, F., {Garnett}, D.~R., {Kennicutt}, R.~C.: {Abundances of Metal-rich
  H II Regions in M51}. \apj \textbf{615},  228--241 (2004)

\bibitem[{{Bresolin} et~al.(2009{\natexlab{a}}){Bresolin}, {Gieren},
  {Kudritzki}, {Pietrzy{\'n}ski}, {Urbaneja} and {Carraro}}]{Bresolin:2009a}
{Bresolin}, F., {Gieren}, W., {Kudritzki}, R., {Pietrzy{\'n}ski}, G.,
  {Urbaneja}, M.~A., {Carraro}, G.: {Extragalactic Chemical Abundances: Do H II
  Regions and Young Stars Tell the Same Story? The Case of the Spiral Galaxy
  NGC 300}. \apj \textbf{700},  309--330 (2009{\natexlab{a}})

\bibitem[{{Bresolin} and {Kennicutt}(2015)}]{Bresolin:2015}
{Bresolin}, F., {Kennicutt}, R.~C.: {Abundance gradients in low surface
  brightness spirals: clues on the origin of common gradients in galactic
  discs}. \mnras \textbf{454},  3664--3673 (2015)

\bibitem[{{Bresolin}, {Kennicutt} and {Ryan-Weber}(2012)}]{Bresolin:2012}
{Bresolin}, F., {Kennicutt}, R.~C., {Ryan-Weber}, E.: {Gas Metallicities in the
  Extended Disks of NGC 1512 and NGC 3621. Chemical Signatures of Metal Mixing
  or Enriched Gas Accretion?} \apj \textbf{750},  122 (2012)

\bibitem[{{Bresolin} and {Kennicutt}(2002)}]{Bresolin:2002}
{Bresolin}, F., {Kennicutt}, R.~C., Jr.: {Optical Spectroscopy of Metal-rich H
  II Regions and Circumnuclear Hot Spots in M83 and NGC 3351}. \apj
  \textbf{572},  838--860 (2002)

\bibitem[{{Bresolin} et~al.(2016){Bresolin}, {Kudritzki}, {Urbaneja}, {Gieren},
  {Ho} and {Pietrzy{\'n}ski}}]{bresolin:2016}
{Bresolin}, F., {Kudritzki}, R.-P., {Urbaneja}, M.~A., {Gieren}, W., {Ho},
  I.-T., {Pietrzy{\'n}ski}, G.: {Young Stars and Ionized Nebulae in M83:
  Comparing Chemical Abundances at High Metallicity.} \apj \textbf{830},  64
  (2016)

\bibitem[{{Bresolin} et~al.(2009{\natexlab{b}}){Bresolin}, {Ryan-Weber},
  {Kennicutt} and {Goddard}}]{Bresolin:2009}
{Bresolin}, F., {Ryan-Weber}, E., {Kennicutt}, R.~C., {Goddard}, Q.: {The Flat
  Oxygen Abundance Gradient in the Extended Disk of M83}. \apj \textbf{695},
  580--595 (2009{\natexlab{b}})

\bibitem[{{Bresolin} et~al.(2005){Bresolin}, {Schaerer}, {Gonz{\'a}lez Delgado}
  and {Stasi{\'n}ska}}]{Bresolin:2005}
{Bresolin}, F., {Schaerer}, D., {Gonz{\'a}lez Delgado}, R.~M., {Stasi{\'n}ska},
  G.: {A VLT study of metal-rich extragalactic H II regions. I. Observations
  and empirical abundances}. \aap \textbf{441},  981--997 (2005)

\bibitem[{{Brown}, {Martini} and {Andrews}(2016)}]{Brown:2016}
{Brown}, J.~S., {Martini}, P., {Andrews}, B.~H.: {A recalibration of
  strong-line oxygen abundance diagnostics via the direct method and
  implications for the high-redshift universe}. \mnras \textbf{458},
  1529--1547 (2016)

\bibitem[{{Carraro} et~al.(2004){Carraro}, {Bresolin}, {Villanova},
  {Matteucci}, {Patat} and {Romaniello}}]{Carraro:2004}
{Carraro}, G., {Bresolin}, F., {Villanova}, S., {Matteucci}, F., {Patat}, F.,
  {Romaniello}, M.: {Metal Abundances in Extremely Distant Galactic Old Open
  Clusters. I. Berkeley 29 and Saurer 1}. \aj \textbf{128},  1676--1683 (2004)

\bibitem[{{Carton} et~al.(2015){Carton}, {Brinchmann}, {Wang}, {Bigiel},
  {Cormier}, {van der Hulst}, {J{\'o}zsa}, {Serra} and
  {Verheijen}}]{Carton:2015}
{Carton}, D., {Brinchmann}, J., {Wang}, J., {Bigiel}, F., {Cormier}, D., {van
  der Hulst}, T., {J{\'o}zsa}, G.~I.~G., {Serra}, P., {Verheijen}, M.~A.~W.:
  {Gas-phase metallicity profiles of the Bluedisk galaxies: Is metallicity in a
  local star formation regulated equilibrium?} \mnras \textbf{451},  210--235
  (2015)

\bibitem[{{Catinella} et~al.(2010){Catinella}, {Schiminovich}, {Kauffmann},
  {Fabello}, {Wang}, {Hummels}, {Lemonias}, {Moran}, {Wu}, {Giovanelli},
  {Haynes}, {Heckman}, {Basu-Zych}, {Blanton}, {Brinchmann}, {Budav{\'a}ri},
  {Gon{\c c}alves}, {Johnson}, {Kennicutt}, {Madore}, {Martin}, {Rich},
  {Tacconi}, {Thilker}, {Wild} and {Wyder}}]{Catinella:2010}
{Catinella}, B., {Schiminovich}, D., {Kauffmann}, G., {Fabello}, S., {Wang},
  J., {Hummels}, C., {Lemonias}, J., {Moran}, S.~M., {Wu}, R., {Giovanelli},
  R., {Haynes}, M.~P., {Heckman}, T.~M., {Basu-Zych}, A.~R., {Blanton}, M.~R.,
  {Brinchmann}, J., {Budav{\'a}ri}, T., {Gon{\c c}alves}, T., {Johnson}, B.~D.,
  {Kennicutt}, R.~C., {Madore}, B.~F., {Martin}, C.~D., {Rich}, M.~R.,
  {Tacconi}, L.~J., {Thilker}, D.~A., {Wild}, V., {Wyder}, T.~K.: {The GALEX
  Arecibo SDSS Survey - I. Gas fraction scaling relations of massive galaxies
  and first data release}. \mnras \textbf{403},  683--708 (2010)

\bibitem[{{Cavichia} et~al.(2014){Cavichia}, {Moll{\'a}}, {Costa} and
  {Maciel}}]{Cavichia:2014}
{Cavichia}, O., {Moll{\'a}}, M., {Costa}, R.~D.~D., {Maciel}, W.~J.: {The role
  of the Galactic bar in the chemical evolution of the Milky Way}. \mnras
  \textbf{437},  3688--3701 (2014)

\bibitem[{{Clarke}(1989)}]{Clarke:1989}
{Clarke}, C.~J.: {Chemical evolution of viscously evolving galactic discs}.
  \mnras \textbf{238},  283--292 (1989)

\bibitem[{{Corradi} et~al.(2015){Corradi}, {Kwitter}, {Balick}, {Henry} and
  {Hensley}}]{Corradi:2015}
{Corradi}, R.~L.~M., {Kwitter}, K.~B., {Balick}, B., {Henry}, R.~B.~C.,
  {Hensley}, K.: {The Chemistry of Planetary Nebulae in the Outer Regions of
  M31}. \apj \textbf{807},  181 (2015)

\bibitem[{{C{\^o}t{\'e}}, {Martel} and {Drissen}(2015)}]{Cote:2015}
{C{\^o}t{\'e}}, B., {Martel}, H., {Drissen}, L.: {Cosmological Simulations of
  the Intergalactic Medium Evolution. II. Galaxy Model and Feedback}. \apj
  \textbf{802},  123 (2015)

\bibitem[{{Croxall} et~al.(2016){Croxall}, {Pogge}, {Berg}, {Skillman} and
  {Moustakas}}]{Croxall:2016}
{Croxall}, K.~V., {Pogge}, R.~W., {Berg}, D.~A., {Skillman}, E.~D.,
  {Moustakas}, J.: {CHAOS III: Gas-phase Abundances in NGC 5457}. \apj
  \textbf{830},  4 (2016)

\bibitem[{{Dav{\'e}}, {Finlator} and {Oppenheimer}(2011)}]{Dave:2011a}
{Dav{\'e}}, R., {Finlator}, K., {Oppenheimer}, B.~D.: {Galaxy evolution in
  cosmological simulations with outflows - II. Metallicities and gas
  fractions}. \mnras \textbf{416},  1354--1376 (2011)

\bibitem[{{de Vaucouleurs} et~al.(1991){de Vaucouleurs}, {de Vaucouleurs},
  {Corwin}, {Buta}, {Paturel} and {Fouque}}]{de-Vaucouleurs:1991}
{de Vaucouleurs}, G., {de Vaucouleurs}, A., {Corwin}, H.~G., Jr., {Buta},
  R.~J., {Paturel}, G., {Fouque}, P.: {Third Reference Catalogue of Bright
  Galaxies}. Springer-Verlag Berlin Heidelberg New York (1991)

\bibitem[{{Di Teodoro} and {Fraternali}(2014)}]{Di-Teodoro:2014}
{Di Teodoro}, E.~M., {Fraternali}, F.: {Gas accretion from minor mergers in
  local spiral galaxies}. \aap \textbf{567},  A68 (2014)

\bibitem[{{Erb} et~al.(2006){Erb}, {Shapley}, {Pettini}, {Steidel}, {Reddy} and
  {Adelberger}}]{Erb:2006}
{Erb}, D.~K., {Shapley}, A.~E., {Pettini}, M., {Steidel}, C.~C., {Reddy},
  N.~A., {Adelberger}, K.~L.: {The Mass-Metallicity Relation at z $>$ 2}. \apj
  \textbf{644},  813--828 (2006)

\bibitem[{{Espada} et~al.(2011)}]{Espada:2011}
{Espada}, D., et~al.: {Star Formation in the Extended Gaseous Disk of the
  Isolated Galaxy CIG 96}. \apj \textbf{736},  20 (2011)

\bibitem[{{Esteban} et~al.(2009){Esteban}, {Bresolin}, {Peimbert},
  {Garc{\'{\i}}a-Rojas}, {Peimbert} and {Mesa-Delgado}}]{Esteban:2009}
{Esteban}, C., {Bresolin}, F., {Peimbert}, M., {Garc{\'{\i}}a-Rojas}, J.,
  {Peimbert}, A., {Mesa-Delgado}, A.: {Keck HIRES Spectroscopy of Extragalactic
  H II Regions: C and O Abundances from Recombination Lines}. \apj
  \textbf{700},  654--678 (2009)

\bibitem[{{Esteban} et~al.(2013){Esteban}, {Carigi}, {Copetti},
  {Garc{\'{\i}}a-Rojas}, {Mesa-Delgado}, {Casta{\~n}eda} and
  {P{\'e}quignot}}]{Esteban:2013}
{Esteban}, C., {Carigi}, L., {Copetti}, M.~V.~F., {Garc{\'{\i}}a-Rojas}, J.,
  {Mesa-Delgado}, A., {Casta{\~n}eda}, H.~O., {P{\'e}quignot}, D.: {NGC 2579
  and the carbon and oxygen abundance gradients beyond the solar circle}.
  \mnras \textbf{433},  382--393 (2013)

\bibitem[{{Ferguson}, {Gallagher} and {Wyse}(1998)}]{Ferguson:1998a}
{Ferguson}, A.~M.~N., {Gallagher}, J.~S., {Wyse}, R.~F.~G.: {The Extreme Outer
  Regions of Disk Galaxies. I. Chemical Abundances of H II Regions}. \aj
  \textbf{116},  673--690 (1998)

\bibitem[{{Finlator} and {Dav{\'e}}(2008)}]{Finlator:2008}
{Finlator}, K., {Dav{\'e}}, R.: {The origin of the galaxy mass-metallicity
  relation and implications for galactic outflows}. \mnras \textbf{385},
  2181--2204 (2008)

\bibitem[{{Freeman}(1970)}]{Freeman:1970}
{Freeman}, K.~C.: {On the Disks of Spiral and S0 Galaxies}. \apj \textbf{160},
  811 (1970)

\bibitem[{{Fu} et~al.(2013){Fu}, {Kauffmann}, {Huang}, {Yates}, {Moran},
  {Heckman}, {Dav{\'e}}, {Guo} and {Henriques}}]{Fu:2013}
{Fu}, J., {Kauffmann}, G., {Huang}, M.-l., {Yates}, R.~M., {Moran}, S.,
  {Heckman}, T.~M., {Dav{\'e}}, R., {Guo}, Q., {Henriques}, B.~M.~B.: {Star
  formation and metallicity gradients in semi-analytic models of disc galaxy
  formation}. \mnras \textbf{434},  1531--1548 (2013)

\bibitem[{{Garc{\'{\i}}a-Rojas} and {Esteban}(2007)}]{Garcia-Rojas:2007a}
{Garc{\'{\i}}a-Rojas}, J., {Esteban}, C.: {On the Abundance Discrepancy Problem
  in H II Regions}. \apj \textbf{670},  457--470 (2007)

\bibitem[{{Garnett}(1990)}]{Garnett:1990}
{Garnett}, D.~R.: {Nitrogen in irregular galaxies}. \apj \textbf{363},
  142--153 (1990)

\bibitem[{{Genovali} et~al.(2015){Genovali}, {Lemasle}, {da Silva}, {Bono},
  {Fabrizio}, {Bergemann}, {Buonanno}, {Ferraro}, {Fran{\c c}ois}, {Iannicola},
  {Inno}, {Laney}, {Kudritzki}, {Matsunaga}, {Nonino}, {Primas}, {Romaniello},
  {Urbaneja} and {Th{\'e}venin}}]{Genovali:2015}
{Genovali}, K., {Lemasle}, B., {da Silva}, R., {Bono}, G., {Fabrizio}, M.,
  {Bergemann}, M., {Buonanno}, R., {Ferraro}, I., {Fran{\c c}ois}, P.,
  {Iannicola}, G., {Inno}, L., {Laney}, C.~D., {Kudritzki}, R.-P., {Matsunaga},
  N., {Nonino}, M., {Primas}, F., {Romaniello}, M., {Urbaneja}, M.~A.,
  {Th{\'e}venin}, F.: {On the {$\alpha$}-element gradients of the Galactic thin
  disk using Cepheids}. \aap \textbf{580},  A17 (2015)

\bibitem[{{Gil de Paz} et~al.(2005){Gil de Paz}, {Madore}, {Boissier},
  {Swaters}, {Popescu}, {Tuffs}, {Sheth}, {Kennicutt}, {Bianchi}, {Thilker} and
  {Martin}}]{Gil-de-Paz:2005}
{Gil de Paz}, A., {Madore}, B.~F., {Boissier}, S., {Swaters}, R., {Popescu},
  C.~C., {Tuffs}, R.~J., {Sheth}, K., {Kennicutt}, R.~C., Jr., {Bianchi}, L.,
  {Thilker}, D., {Martin}, D.~C.: {Discovery of an Extended Ultraviolet Disk in
  the Nearby Galaxy NGC 4625}. \apjl \textbf{627},  L29--L32 (2005)

\bibitem[{{Gil de Paz} et~al.(2007){Gil de Paz}, {Madore}, {Boissier},
  {Thilker}, {Bianchi}, {S{\'a}nchez Contreras}, {Barlow}, {Conrow}, {Forster},
  {Friedman}, {Martin}, {Morrissey}, {Neff}, {Rich}, {Schiminovich}, {Seibert},
  {Small}, {Donas}, {Heckman}, {Lee}, {Milliard}, {Szalay}, {Wyder} and
  {Yi}}]{Gil-de-Paz:2007}
{Gil de Paz}, A., {Madore}, B.~F., {Boissier}, S., {Thilker}, D., {Bianchi},
  L., {S{\'a}nchez Contreras}, C., {Barlow}, T.~A., {Conrow}, T., {Forster},
  K., {Friedman}, P.~G., {Martin}, D.~C., {Morrissey}, P., {Neff}, S.~G.,
  {Rich}, R.~M., {Schiminovich}, D., {Seibert}, M., {Small}, T., {Donas}, J.,
  {Heckman}, T.~M., {Lee}, Y.-W., {Milliard}, B., {Szalay}, A.~S., {Wyder},
  T.~K., {Yi}, S.: {Chemical and Photometric Evolution of Extended Ultraviolet
  Disks: Optical Spectroscopy of M83 (NGC 5236) and NGC 4625}. \apj
  \textbf{661},  115--134 (2007)

\bibitem[{{Goddard} et~al.(2011){Goddard}, {Bresolin}, {Kennicutt},
  {Ryan-Weber} and {Rosales-Ortega}}]{Goddard:2011}
{Goddard}, Q.~E., {Bresolin}, F., {Kennicutt}, R.~C., {Ryan-Weber}, E.~V.,
  {Rosales-Ortega}, F.~F.: {On the nature of the H II regions in the extended
  ultraviolet disc of NGC 4625}. \mnras \textbf{412},  1246--1258 (2011)

\bibitem[{{Goddard}, {Kennicutt} and {Ryan-Weber}(2010)}]{Goddard:2010}
{Goddard}, Q.~E., {Kennicutt}, R.~C., {Ryan-Weber}, E.~V.: {On the nature of
  star formation at large galactic radii}. \mnras \textbf{405},  2791--2809
  (2010)

\bibitem[{{Graham} and {Driver}(2005)}]{Graham:2005}
{Graham}, A.~W., {Driver}, S.~P.: {A Concise Reference to (Projected)
  S{\'e}rsic R$^{1/n}$ Quantities, Including Concentration, Profile Slopes,
  Petrosian Indices, and Kron Magnitudes}. \pasa \textbf{22},  118--127 (2005)

\bibitem[{{Henry} and {Worthey}(1999)}]{Henry:1999}
{Henry}, R.~B.~C., {Worthey}, G.: {The Distribution of Heavy Elements in Spiral
  and Elliptical Galaxies}. \pasp \textbf{111},  919--945 (1999)

\bibitem[{{Ho} et~al.(2015){Ho}, {Kudritzki}, {Kewley}, {Zahid}, {Dopita},
  {Bresolin} and {Rupke}}]{Ho:2015}
{Ho}, I.-T., {Kudritzki}, R.-P., {Kewley}, L.~J., {Zahid}, H.~J., {Dopita},
  M.~A., {Bresolin}, F., {Rupke}, D.~S.~N.: {Metallicity gradients in local
  field star-forming galaxies: insights on inflows, outflows, and the
  coevolution of gas, stars and metals}. \mnras \textbf{448},  2030--2054
  (2015)

\bibitem[{{Kennicutt}, {Bresolin} and {Garnett}(2003)}]{Kennicutt:2003}
{Kennicutt}, R.~C., {Bresolin}, F., {Garnett}, D.~R.: {The Composition Gradient
  in M101 Revisited. II. Electron Temperatures and Implications for the Nebular
  Abundance Scale}. \apj \textbf{591},  801--820 (2003)

\bibitem[{{Kewley} and {Dopita}(2002)}]{Kewley:2002}
{Kewley}, L.~J., {Dopita}, M.~A.: {Using Strong Lines to Estimate Abundances in
  Extragalactic H II Regions and Starburst Galaxies}. \apjs \textbf{142},
  35--52 (2002)

\bibitem[{{Kewley} and {Ellison}(2008)}]{Kewley:2008}
{Kewley}, L.~J., {Ellison}, S.~L.: {Metallicity Calibrations and the
  Mass-Metallicity Relation for Star-forming Galaxies}. \apj \textbf{681},
  1183--1204 (2008)

\bibitem[{{Kobayashi}, {Springel} and {White}(2007)}]{Kobayashi:2007}
{Kobayashi}, C., {Springel}, V., {White}, S.~D.~M.: {Simulations of Cosmic
  Chemical Enrichment}. \mnras \textbf{376},  1465--1479 (2007)

\bibitem[{{Kobayashi} et~al.(2006){Kobayashi}, {Umeda}, {Nomoto}, {Tominaga}
  and {Ohkubo}}]{Kobayashi:2006}
{Kobayashi}, C., {Umeda}, H., {Nomoto}, K., {Tominaga}, N., {Ohkubo}, T.:
  {Galactic Chemical Evolution: Carbon through Zinc}. \apj \textbf{653},
  1145--1171 (2006)

\bibitem[{{Kobulnicky} and {Kewley}(2004)}]{Kobulnicky:2004}
{Kobulnicky}, H.~A., {Kewley}, L.~J.: {Metallicities of 0.3$<$z$<$1.0 Galaxies
  in the GOODS-North Field}. \apj \textbf{617},  240--261 (2004)

\bibitem[{{Koribalski} and {L{\'o}pez-S{\'a}nchez}(2009)}]{Koribalski:2009}
{Koribalski}, B.~S., {L{\'o}pez-S{\'a}nchez}, {\'A}.~R.: {Gas dynamics and star
  formation in the galaxy pair NGC1512/1510}. \mnras \textbf{400},  1749--1767
  (2009)

\bibitem[{{Korotin} et~al.(2014){Korotin}, {Andrievsky}, {Luck}, {L{\'e}pine},
  {Maciel} and {Kovtyukh}}]{Korotin:2014}
{Korotin}, S.~A., {Andrievsky}, S.~M., {Luck}, R.~E., {L{\'e}pine}, J.~R.~D.,
  {Maciel}, W.~J., {Kovtyukh}, V.~V.: {Oxygen abundance distribution in the
  Galactic disc}. \mnras \textbf{444},  3301--3307 (2014)

\bibitem[{{Krumholz} and {Burkhart}(2016)}]{Krumholz:2016}
{Krumholz}, M.~R., {Burkhart}, B.: {Is turbulence in the interstellar medium
  driven by feedback or gravity? An observational test}. \mnras \textbf{458},
  1671--1677 (2016)

\bibitem[{{Kubryk}, {Prantzos} and {Athanassoula}(2015)}]{Kubryk:2015}
{Kubryk}, M., {Prantzos}, N., {Athanassoula}, E.: {Evolution of the Milky Way
  with radial motions of stars and gas. II. The evolution of abundance profiles
  from H to Ni}. \aap \textbf{580},  A127 (2015)

\bibitem[{{Kudritzki} et~al.(2015){Kudritzki}, {Ho}, {Schruba}, {Burkert},
  {Zahid}, {Bresolin} and {Dima}}]{Kudritzki:2015}
{Kudritzki}, R.-P., {Ho}, I.-T., {Schruba}, A., {Burkert}, A., {Zahid}, H.~J.,
  {Bresolin}, F., {Dima}, G.~I.: {The chemical evolution of local star-forming
  galaxies: radial profiles of ISM metallicity, gas mass, and stellar mass and
  constraints on galactic accretion and winds}. \mnras \textbf{450},  342--359
  (2015)

\bibitem[{{Kudritzki} et~al.(2014){Kudritzki}, {Urbaneja}, {Bresolin}, {Hosek}
  and {Przybilla}}]{Kudritzki:2014}
{Kudritzki}, R.-P., {Urbaneja}, M.~A., {Bresolin}, F., {Hosek}, M.~W., Jr.,
  {Przybilla}, N.: {Stellar Metallicity of the Extended Disk and Distance of
  the Spiral Galaxy NGC 3621}. \apj \textbf{788},  56 (2014)

\bibitem[{{Lacey} and {Fall}(1985)}]{Lacey:1985}
{Lacey}, C.~G., {Fall}, S.~M.: {Chemical evolution of the galactic disk with
  radial gas flows}. \apj \textbf{290},  154--170 (1985)

\bibitem[{{Lehner} et~al.(2013){Lehner}, {Howk}, {Tripp}, {Tumlinson},
  {Prochaska}, {O'Meara}, {Thom}, {Werk}, {Fox} and {Ribaudo}}]{Lehner:2013}
{Lehner}, N., {Howk}, J.~C., {Tripp}, T.~M., {Tumlinson}, J., {Prochaska},
  J.~X., {O'Meara}, J.~M., {Thom}, C., {Werk}, J.~K., {Fox}, A.~J., {Ribaudo},
  J.: {The Bimodal Metallicity Distribution of the Cool Circumgalactic Medium
  at z $<$ 1}. \apj \textbf{770},  138 (2013)

\bibitem[{{Lehnert}, {van Driel} and {Minchin}(2016)}]{Lehnert:2016}
{Lehnert}, M.~D., {van Driel}, W., {Minchin}, R.: {Can galaxy growth be
  sustained through HI-rich minor mergers?} \aap \textbf{590},  A51 (2016)

\bibitem[{{Lemasle} et~al.(2013){Lemasle}, {Fran{\c c}ois}, {Genovali},
  {Kovtyukh}, {Bono}, {Inno}, {Laney}, {Kaper}, {Bergemann}, {Fabrizio},
  {Matsunaga}, {Pedicelli}, {Primas} and {Romaniello}}]{Lemasle:2013}
{Lemasle}, B., {Fran{\c c}ois}, P., {Genovali}, K., {Kovtyukh}, V.~V., {Bono},
  G., {Inno}, L., {Laney}, C.~D., {Kaper}, L., {Bergemann}, M., {Fabrizio}, M.,
  {Matsunaga}, N., {Pedicelli}, S., {Primas}, F., {Romaniello}, M.: {Galactic
  abundance gradients from Cepheids. {$\alpha$} and heavy elements in the outer
  disk}. \aap \textbf{558},  A31 (2013)

\bibitem[{{L{\'e}pine} et~al.(2014){L{\'e}pine}, {Andrievky}, {Barros},
  {Junqueira} and {Scarano}}]{Lepine:2014}
{L{\'e}pine}, J.~R.~D., {Andrievky}, S., {Barros}, D.~A., {Junqueira}, T.~C.,
  {Scarano}, S.: {Bimodal chemical evolution of the Galactic disk and the
  Barium abundance of Cepheids}. In S.~{Feltzing}, G.~{Zhao}, N.~A. {Walton},
  P.~{Whitelock}, eds., IAU Symposium (2014), volume 298 of IAU Symposium,
  86--91

\bibitem[{{L{\'e}pine} et~al.(2011){L{\'e}pine}, {Cruz}, {Scarano}, {Barros},
  {Dias}, {Pomp{\'e}ia}, {Andrievsky}, {Carraro} and {Famaey}}]{Lepine:2011}
{L{\'e}pine}, J.~R.~D., {Cruz}, P., {Scarano}, S., Jr., {Barros}, D.~A.,
  {Dias}, W.~S., {Pomp{\'e}ia}, L., {Andrievsky}, S.~M., {Carraro}, G.,
  {Famaey}, B.: {Overlapping abundance gradients and azimuthal gradients
  related to the spiral structure of the Galaxy}. \mnras \textbf{417},
  698--708 (2011)

\bibitem[{{Lilly} et~al.(2013){Lilly}, {Carollo}, {Pipino}, {Renzini} and
  {Peng}}]{Lilly:2013}
{Lilly}, S.~J., {Carollo}, C.~M., {Pipino}, A., {Renzini}, A., {Peng}, Y.: {Gas
  Regulation of Galaxies: The Evolution of the Cosmic Specific Star Formation
  Rate, the Metallicity-Mass-Star-formation Rate Relation, and the Stellar
  Content of Halos}. \apj \textbf{772},  119 (2013)

\bibitem[{{L{\'o}pez-S{\'a}nchez} et~al.(2012){L{\'o}pez-S{\'a}nchez},
  {Dopita}, {Kewley}, {Zahid}, {Nicholls} and
  {Scharw{\"a}chter}}]{Lopez-Sanchez:2012}
{L{\'o}pez-S{\'a}nchez}, {\'A}.~R., {Dopita}, M.~A., {Kewley}, L.~J., {Zahid},
  H.~J., {Nicholls}, D.~C., {Scharw{\"a}chter}, J.: {Eliminating error in the
  chemical abundance scale for extragalactic H II regions}. \mnras
  \textbf{426},  2630--2651 (2012)

\bibitem[{{L{\'o}pez-S{\'a}nchez} et~al.(2015){L{\'o}pez-S{\'a}nchez},
  {Westmeier}, {Esteban} and {Koribalski}}]{Lopez-Sanchez:2015}
{L{\'o}pez-S{\'a}nchez}, {\'A}.~R., {Westmeier}, T., {Esteban}, C.,
  {Koribalski}, B.~S.: {Ionized gas in the XUV disc of the NGC 1512/1510
  system}. \mnras \textbf{450},  3381--3409 (2015)

\bibitem[{{Magrini} et~al.(2016){Magrini}, {Coccato}, {Stanghellini},
  {Casasola} and {Galli}}]{Magrini:2016}
{Magrini}, L., {Coccato}, L., {Stanghellini}, L., {Casasola}, V., {Galli}, D.:
  {Metallicity gradients in local Universe galaxies: Time evolution and effects
  of radial migration}. \aap \textbf{588},  A91 (2016)

\bibitem[{{Magrini} et~al.(2009){Magrini}, {Sestito}, {Randich} and
  {Galli}}]{Magrini:2009b}
{Magrini}, L., {Sestito}, P., {Randich}, S., {Galli}, D.: {The evolution of the
  Galactic metallicity gradient from high-resolution spectroscopy of open
  clusters}. \aap \textbf{494},  95--108 (2009)

\bibitem[{{Maiolino} et~al.(2008){Maiolino}, {Nagao}, {Grazian}, {Cocchia},
  {Marconi}, {Mannucci}, {Cimatti}, {Pipino}, {Ballero}, {Calura}, {Chiappini},
  {Fontana}, {Granato}, {Matteucci}, {Pastorini}, {Pentericci}, {Risaliti},
  {Salvati} and {Silva}}]{Maiolino:2008}
{Maiolino}, R., {Nagao}, T., {Grazian}, A., {Cocchia}, F., {Marconi}, A.,
  {Mannucci}, F., {Cimatti}, A., {Pipino}, A., {Ballero}, S., {Calura}, F.,
  {Chiappini}, C., {Fontana}, A., {Granato}, G.~L., {Matteucci}, F.,
  {Pastorini}, G., {Pentericci}, L., {Risaliti}, G., {Salvati}, M., {Silva},
  L.: {AMAZE. I. The evolution of the mass-metallicity relation at z $>$ 3}.
  \aap \textbf{488},  463--479 (2008)

\bibitem[{{Mannucci} et~al.(2010){Mannucci}, {Cresci}, {Maiolino}, {Marconi}
  and {Gnerucci}}]{Mannucci:2010}
{Mannucci}, F., {Cresci}, G., {Maiolino}, R., {Marconi}, A., {Gnerucci}, A.: {A
  fundamental relation between mass, star formation rate and metallicity in
  local and high-redshift galaxies}. \mnras \textbf{408},  2115--2127 (2010)

\bibitem[{{Marino} et~al.(2012){Marino}, {Gil de Paz}, {Castillo-Morales},
  {Mu{\~n}oz-Mateos}, {S{\'a}nchez}, {P{\'e}rez-Gonz{\'a}lez}, {Gallego},
  {Zamorano}, {Alonso-Herrero} and {Boissier}}]{Marino:2012}
{Marino}, R.~A., {Gil de Paz}, A., {Castillo-Morales}, A., {Mu{\~n}oz-Mateos},
  J.~C., {S{\'a}nchez}, S.~F., {P{\'e}rez-Gonz{\'a}lez}, P.~G., {Gallego}, J.,
  {Zamorano}, J., {Alonso-Herrero}, A., {Boissier}, S.: {Integral Field
  Spectroscopy and Multi-wavelength Imaging of the nearby Spiral Galaxy NGC
  5668: An Unusual Flattening in Metallicity Gradient}. \apj \textbf{754},  61
  (2012)

\bibitem[{{Marino} et~al.(2016){Marino}, {Gil de Paz}, {S{\'a}nchez},
  {S{\'a}nchez-Bl{\'a}zquez}, {Cardiel}, {Castillo-Morales}, {Pascual},
  {V{\'{\i}}lchez}, {Kehrig}, {Moll{\'a}}, {Mendez-Abreu},
  {Catal{\'a}n-Torrecilla}, {Florido}, {Perez}, {Ruiz-Lara}, {Ellis},
  {L{\'o}pez-S{\'a}nchez}, {Gonz{\'a}lez Delgado}, {de Lorenzo-C{\'a}ceres},
  {Garc{\'{\i}}a-Benito}, {Galbany}, {Zibetti}, {Cortijo}, {Kalinova}, {Mast},
  {Iglesias-P{\'a}ramo}, {Papaderos}, {Walcher} and
  {Bland-Hawthorn}}]{Marino:2016}
{Marino}, R.~A., {Gil de Paz}, A., {S{\'a}nchez}, S.~F.,
  {S{\'a}nchez-Bl{\'a}zquez}, P., {Cardiel}, N., {Castillo-Morales}, A.,
  {Pascual}, S., {V{\'{\i}}lchez}, J., {Kehrig}, C., {Moll{\'a}}, M.,
  {Mendez-Abreu}, J., {Catal{\'a}n-Torrecilla}, C., {Florido}, E., {Perez}, I.,
  {Ruiz-Lara}, T., {Ellis}, S., {L{\'o}pez-S{\'a}nchez}, A.~R., {Gonz{\'a}lez
  Delgado}, R.~M., {de Lorenzo-C{\'a}ceres}, A., {Garc{\'{\i}}a-Benito}, R.,
  {Galbany}, L., {Zibetti}, S., {Cortijo}, C., {Kalinova}, V., {Mast}, D.,
  {Iglesias-P{\'a}ramo}, J., {Papaderos}, P., {Walcher}, C.~J.,
  {Bland-Hawthorn}, J.: {Outer-disk reddening and gas-phase metallicities: The
  CALIFA connection}. \aap \textbf{585},  A47 (2016)

\bibitem[{{Marino} et~al.(2013){Marino}, {Rosales-Ortega}, {S{\'a}nchez}, {Gil
  de Paz}, {V{\'{\i}}lchez}, {Miralles-Caballero}, {Kehrig},
  {P{\'e}rez-Montero}, {Stanishev}, {Iglesias-P{\'a}ramo}, {D{\'{\i}}az},
  {Castillo-Morales}, {Kennicutt}, {L{\'o}pez-S{\'a}nchez}, {Galbany},
  {Garc{\'{\i}}a-Benito}, {Mast}, {Mendez-Abreu}, {Monreal-Ibero}, {Husemann},
  {Walcher}, {Garc{\'{\i}}a-Lorenzo}, {Masegosa}, {Del Olmo Orozco},
  {Mour{\~a}o}, {Ziegler}, {Moll{\'a}}, {Papaderos},
  {S{\'a}nchez-Bl{\'a}zquez}, {Gonz{\'a}lez Delgado}, {Falc{\'o}n-Barroso},
  {Roth}, {van de Ven} and {Califa Team}}]{Marino:2013}
{Marino}, R.~A., {Rosales-Ortega}, F.~F., {S{\'a}nchez}, S.~F., {Gil de Paz},
  A., {V{\'{\i}}lchez}, J., {Miralles-Caballero}, D., {Kehrig}, C.,
  {P{\'e}rez-Montero}, E., {Stanishev}, V., {Iglesias-P{\'a}ramo}, J.,
  {D{\'{\i}}az}, A.~I., {Castillo-Morales}, A., {Kennicutt}, R.,
  {L{\'o}pez-S{\'a}nchez}, A.~R., {Galbany}, L., {Garc{\'{\i}}a-Benito}, R.,
  {Mast}, D., {Mendez-Abreu}, J., {Monreal-Ibero}, A., {Husemann}, B.,
  {Walcher}, C.~J., {Garc{\'{\i}}a-Lorenzo}, B., {Masegosa}, J., {Del Olmo
  Orozco}, A., {Mour{\~a}o}, A.~M., {Ziegler}, B., {Moll{\'a}}, M.,
  {Papaderos}, P., {S{\'a}nchez-Bl{\'a}zquez}, P., {Gonz{\'a}lez Delgado},
  R.~M., {Falc{\'o}n-Barroso}, J., {Roth}, M.~M., {van de Ven}, G., {Califa
  Team}: {The O3N2 and N2 abundance indicators revisited: improved calibrations
  based on CALIFA and T$_{e}$-based literature data}. \aap \textbf{559},  A114
  (2013)

\bibitem[{{Martin} and {Roy}(1995)}]{Martin:1995}
{Martin}, P., {Roy}, J.-R.: {The oxygen distribution in NGC 3359 or a disk
  galaxy in the early phase of bar formation}. \apj \textbf{445},  161--172
  (1995)

\bibitem[{{Matteucci} and {Chiosi}(1983)}]{Matteucci:1983}
{Matteucci}, F., {Chiosi}, C.: {Stochastic star formation and chemical
  evolution of dwarf irregular galaxies}. \aap \textbf{123},  121--134 (1983)

\bibitem[{{McGaugh}(1991)}]{McGaugh:1991}
{McGaugh}, S.~S.: {H II region abundances - Model oxygen line ratios}. \apj
  \textbf{380},  140--150 (1991)

\bibitem[{{Menzel}, {Aller} and {Hebb}(1941)}]{Menzel:1941}
{Menzel}, D.~H., {Aller}, L.~H., {Hebb}, M.~H.: {Physical Processes in Gaseous
  Nebulae. XIII.} \apj \textbf{93},  230 (1941)

\bibitem[{{Minchev} et~al.(2011){Minchev}, {Famaey}, {Combes}, {Di Matteo},
  {Mouhcine} and {Wozniak}}]{Minchev:2011}
{Minchev}, I., {Famaey}, B., {Combes}, F., {Di Matteo}, P., {Mouhcine}, M.,
  {Wozniak}, H.: {Radial migration in galactic disks caused by resonance
  overlap of multiple patterns: Self-consistent simulations}. \aap
  \textbf{527},  A147 (2011)

\bibitem[{{Mishurov}, {L{\'e}pine} and {Acharova}(2002)}]{Mishurov:2002}
{Mishurov}, Y.~N., {L{\'e}pine}, J.~R.~D., {Acharova}, I.~A.: {Corotation: Its
  Influence on the Chemical Abundance Pattern of the Galaxy}. \apjl
  \textbf{571},  L113--L115 (2002)

\bibitem[{{Moran} et~al.(2012){Moran}, {Heckman}, {Kauffmann}, {Dav{\'e}},
  {Catinella}, {Brinchmann}, {Wang}, {Schiminovich}, {Saintonge},
  {Gracia-Carpio}, {Tacconi}, {Giovanelli}, {Haynes}, {Fabello}, {Hummels},
  {Lemonias} and {Wu}}]{Moran:2012}
{Moran}, S.~M., {Heckman}, T.~M., {Kauffmann}, G., {Dav{\'e}}, R., {Catinella},
  B., {Brinchmann}, J., {Wang}, J., {Schiminovich}, D., {Saintonge}, A.,
  {Gracia-Carpio}, J., {Tacconi}, L., {Giovanelli}, R., {Haynes}, M.,
  {Fabello}, S., {Hummels}, C., {Lemonias}, J., {Wu}, R.: {The GALEX Arecibo
  SDSS Survey. V. The Relation between the H I Content of Galaxies and Metal
  Enrichment at Their Outskirts}. \apj \textbf{745},  66 (2012)

\bibitem[{{Olave-Rojas} et~al.(2015){Olave-Rojas}, {Torres-Flores}, {Carrasco},
  {Mendes de Oliveira}, {de Mello} and {Scarano}}]{Olave-Rojas:2015}
{Olave-Rojas}, D., {Torres-Flores}, S., {Carrasco}, E.~R., {Mendes de
  Oliveira}, C., {de Mello}, D.~F., {Scarano}, S.: {NGC 6845: metallicity
  gradients and star formation in a complex compact group}. \mnras
  \textbf{453},  2808--2823 (2015)

\bibitem[{{Oppenheimer} and {Dav{\'e}}(2008)}]{Oppenheimer:2008}
{Oppenheimer}, B.~D., {Dav{\'e}}, R.: {Mass, metal, and energy feedback in
  cosmological simulations}. \mnras \textbf{387},  577--600 (2008)

\bibitem[{{Pagel} et~al.(1979){Pagel}, {Edmunds}, {Blackwell}, {Chun} and
  {Smith}}]{Pagel:1979}
{Pagel}, B.~E.~J., {Edmunds}, M.~G., {Blackwell}, D.~E., {Chun}, M.~S.,
  {Smith}, G.: {On the composition of H II regions in southern galaxies. I -
  NGC 300 and 1365}. \mnras \textbf{189},  95--113 (1979)

\bibitem[{{Patterson} et~al.(2012){Patterson}, {Walterbos}, {Kennicutt},
  {Chiappini} and {Thilker}}]{Patterson:2012}
{Patterson}, M.~T., {Walterbos}, R.~A.~M., {Kennicutt}, R.~C., {Chiappini}, C.,
  {Thilker}, D.~A.: {An oxygen abundance gradient into the outer disc of M81}.
  \mnras \textbf{422},  401--419 (2012)

\bibitem[{{Petit} et~al.(2015){Petit}, {Krumholz}, {Goldbaum} and
  {Forbes}}]{Petit:2015}
{Petit}, A.~C., {Krumholz}, M.~R., {Goldbaum}, N.~J., {Forbes}, J.~C.: {Mixing
  and transport of metals by gravitational instability-driven turbulence in
  galactic discs}. \mnras \textbf{449},  2588--2597 (2015)

\bibitem[{{Pettini} and {Pagel}(2004)}]{Pettini:2004}
{Pettini}, M., {Pagel}, B.~E.~J.: {[OIII]/[NII] as an abundance indicator at
  high redshift}. \mnras \textbf{348},  L59--L63 (2004)

\bibitem[{{Pilyugin} and {Grebel}(2016)}]{Pilyugin:2016}
{Pilyugin}, L.~S., {Grebel}, E.~K.: {New calibrations for abundance
  determinations in H II regions}. \mnras \textbf{457},  3678--3692 (2016)

\bibitem[{{Pilyugin}, {Grebel} and {Mattsson}(2012)}]{Pilyugin:2012a}
{Pilyugin}, L.~S., {Grebel}, E.~K., {Mattsson}, L.: {'Counterpart' method for
  abundance determinations in H II regions}. \mnras \textbf{424},  2316--2329
  (2012)

\bibitem[{{Pilyugin} and {Thuan}(2005)}]{Pilyugin:2005a}
{Pilyugin}, L.~S., {Thuan}, T.~X.: {Oxygen Abundance Determination in H II
  Regions: The Strong Line Intensities-Abundance Calibration Revisited}. \apj
  \textbf{631},  231--243 (2005)

\bibitem[{{Rich} et~al.(2012){Rich}, {Torrey}, {Kewley}, {Dopita} and
  {Rupke}}]{Rich:2012}
{Rich}, J.~A., {Torrey}, P., {Kewley}, L.~J., {Dopita}, M.~A., {Rupke},
  D.~S.~N.: {An Integral Field Study of Abundance Gradients in nearby Luminous
  Infrared Galaxies}. \apj \textbf{753},  5 (2012)

\bibitem[{{Rosales-Ortega} et~al.(2011){Rosales-Ortega}, {D{\'{\i}}az},
  {Kennicutt} and {S{\'a}nchez}}]{Rosales-Ortega:2011}
{Rosales-Ortega}, F.~F., {D{\'{\i}}az}, A.~I., {Kennicutt}, R.~C.,
  {S{\'a}nchez}, S.~F.: {PPAK wide-field Integral Field Spectroscopy of NGC 628
  - II. Emission line abundance analysis}. \mnras \textbf{415},  2439--2474
  (2011)

\bibitem[{{Ro{\v s}kar} et~al.(2008){Ro{\v s}kar}, {Debattista}, {Quinn},
  {Stinson} and {Wadsley}}]{Roskar:2008b}
{Ro{\v s}kar}, R., {Debattista}, V.~P., {Quinn}, T.~R., {Stinson}, G.~S.,
  {Wadsley}, J.: {Riding the Spiral Waves: Implications of Stellar Migration
  for the Properties of Galactic Disks}. \apjl \textbf{684},  L79--L82 (2008)

\bibitem[{{Rupke}, {Kewley} and {Barnes}(2010)}]{Rupke:2010}
{Rupke}, D.~S.~N., {Kewley}, L.~J., {Barnes}, J.~E.: {Galaxy Mergers and the
  Mass-Metallicity Relation: Evidence for Nuclear Metal Dilution and Flattened
  Gradients from Numerical Simulations}. \apjl \textbf{710},  L156--L160 (2010)

\bibitem[{{Rupke}, {Kewley} and {Chien}(2010)}]{Rupke:2010a}
{Rupke}, D.~S.~N., {Kewley}, L.~J., {Chien}, L.-H.: {Gas-phase Oxygen Gradients
  in Strongly Interacting Galaxies. I. Early-stage Interactions}. \apj
  \textbf{723},  1255--1271 (2010)

\bibitem[{{S{\'a}nchez} et~al.(2014){S{\'a}nchez}, {Rosales-Ortega},
  {Iglesias-P{\'a}ramo}, {Moll{\'a}}, {Barrera-Ballesteros}, {Marino},
  {P{\'e}rez}, {S{\'a}nchez-Blazquez}, {Gonz{\'a}lez Delgado}, {Cid Fernandes},
  {de Lorenzo-C{\'a}ceres}, {Mendez-Abreu}, {Galbany}, {Falcon-Barroso},
  {Miralles-Caballero}, {Husemann}, {Garc{\'{\i}}a-Benito}, {Mast}, {Walcher},
  {Gil de Paz}, {Garc{\'{\i}}a-Lorenzo}, {Jungwiert}, {V{\'{\i}}lchez},
  {J{\'{\i}}lkov{\'a}}, {Lyubenova}, {Cortijo-Ferrero}, {D{\'{\i}}az},
  {Wisotzki}, {M{\'a}rquez}, {Bland-Hawthorn}, {Ellis}, {van de Ven}, {Jahnke},
  {Papaderos}, {Gomes}, {Mendoza} and {L{\'o}pez-S{\'a}nchez}}]{Sanchez:2014}
{S{\'a}nchez}, S.~F., {Rosales-Ortega}, F.~F., {Iglesias-P{\'a}ramo}, J.,
  {Moll{\'a}}, M., {Barrera-Ballesteros}, J., {Marino}, R.~A., {P{\'e}rez}, E.,
  {S{\'a}nchez-Blazquez}, P., {Gonz{\'a}lez Delgado}, R., {Cid Fernandes}, R.,
  {de Lorenzo-C{\'a}ceres}, A., {Mendez-Abreu}, J., {Galbany}, L.,
  {Falcon-Barroso}, J., {Miralles-Caballero}, D., {Husemann}, B.,
  {Garc{\'{\i}}a-Benito}, R., {Mast}, D., {Walcher}, C.~J., {Gil de Paz}, A.,
  {Garc{\'{\i}}a-Lorenzo}, B., {Jungwiert}, B., {V{\'{\i}}lchez}, J.~M.,
  {J{\'{\i}}lkov{\'a}}, L., {Lyubenova}, M., {Cortijo-Ferrero}, C.,
  {D{\'{\i}}az}, A.~I., {Wisotzki}, L., {M{\'a}rquez}, I., {Bland-Hawthorn},
  J., {Ellis}, S., {van de Ven}, G., {Jahnke}, K., {Papaderos}, P., {Gomes},
  J.~M., {Mendoza}, M.~A., {L{\'o}pez-S{\'a}nchez}, {\'A}.~R.: {A
  characteristic oxygen abundance gradient in galaxy disks unveiled with
  CALIFA}. \aap \textbf{563},  A49 (2014)

\bibitem[{{S{\'a}nchez} et~al.(2012){S{\'a}nchez}, {Rosales-Ortega}, {Marino},
  {Iglesias-P{\'a}ramo}, {V{\'{\i}}lchez}, {Kennicutt}, {D{\'{\i}}az}, {Mast},
  {Monreal-Ibero}, {Garc{\'{\i}}a-Benito}, {Bland-Hawthorn}, {P{\'e}rez},
  {Gonz{\'a}lez Delgado}, {Husemann}, {L{\'o}pez-S{\'a}nchez}, {Cid Fernandes},
  {Kehrig}, {Walcher}, {Gil de Paz} and {Ellis}}]{Sanchez:2012}
{S{\'a}nchez}, S.~F., {Rosales-Ortega}, F.~F., {Marino}, R.~A.,
  {Iglesias-P{\'a}ramo}, J., {V{\'{\i}}lchez}, J.~M., {Kennicutt}, R.~C.,
  {D{\'{\i}}az}, A.~I., {Mast}, D., {Monreal-Ibero}, A.,
  {Garc{\'{\i}}a-Benito}, R., {Bland-Hawthorn}, J., {P{\'e}rez}, E.,
  {Gonz{\'a}lez Delgado}, R., {Husemann}, B., {L{\'o}pez-S{\'a}nchez},
  {\'A}.~R., {Cid Fernandes}, R., {Kehrig}, C., {Walcher}, C.~J., {Gil de Paz},
  A., {Ellis}, S.: {Integral field spectroscopy of a sample of nearby galaxies.
  II. Properties of the H ii regions}. \aap \textbf{546},  A2 (2012)

\bibitem[{{S{\'a}nchez-Menguiano} et~al.(2016){S{\'a}nchez-Menguiano},
  {S{\'a}nchez}, {P{\'e}rez}, {Garc{\'{\i}}a-Benito}, {Husemann}, {Mast},
  {Mendoza}, {Ruiz-Lara}, {Ascasibar}, {Bland-Hawthorn}, {Cavichia},
  {D{\'{\i}}az}, {Florido}, {Galbany}, {G{\'o}nzalez Delgado}, {Kehrig},
  {Marino}, {M{\'a}rquez}, {Masegosa}, {M{\'e}ndez-Abreu}, {Moll{\'a}}, {Del
  Olmo}, {P{\'e}rez}, {S{\'a}nchez-Bl{\'a}zquez}, {Stanishev}, {Walcher},
  {L{\'o}pez-S{\'a}nchez} and {Califa Collaboration}}]{Sanchez-Menguiano:2016}
{S{\'a}nchez-Menguiano}, L., {S{\'a}nchez}, S.~F., {P{\'e}rez}, I.,
  {Garc{\'{\i}}a-Benito}, R., {Husemann}, B., {Mast}, D., {Mendoza}, A.,
  {Ruiz-Lara}, T., {Ascasibar}, Y., {Bland-Hawthorn}, J., {Cavichia}, O.,
  {D{\'{\i}}az}, A.~I., {Florido}, E., {Galbany}, L., {G{\'o}nzalez Delgado},
  R.~M., {Kehrig}, C., {Marino}, R.~A., {M{\'a}rquez}, I., {Masegosa}, J.,
  {M{\'e}ndez-Abreu}, J., {Moll{\'a}}, M., {Del Olmo}, A., {P{\'e}rez}, E.,
  {S{\'a}nchez-Bl{\'a}zquez}, P., {Stanishev}, V., {Walcher}, C.~J.,
  {L{\'o}pez-S{\'a}nchez}, {\'A}.~R., {Califa Collaboration}: {Shape of the
  oxygen abundance profiles in CALIFA face-on spiral galaxies}. \aap
  \textbf{587},  A70 (2016)

\bibitem[{{Sancisi} et~al.(2008){Sancisi}, {Fraternali}, {Oosterloo} and {van
  der Hulst}}]{Sancisi:2008}
{Sancisi}, R., {Fraternali}, F., {Oosterloo}, T., {van der Hulst}, T.: {Cold
  gas accretion in galaxies}. \aapr \textbf{15},  189--223 (2008)

\bibitem[{{Sanders} et~al.(2012){Sanders}, {Caldwell}, {McDowell} and
  {Harding}}]{Sanders:2012}
{Sanders}, N.~E., {Caldwell}, N., {McDowell}, J., {Harding}, P.: {The
  Metallicity Profile of M31 from Spectroscopy of Hundreds of H II Regions and
  PNe}. \apj \textbf{758},  133 (2012)

\bibitem[{{Sanders} et~al.(2015){Sanders}, {Shapley}, {Kriek}, {Reddy},
  {Freeman}, {Coil}, {Siana}, {Mobasher}, {Shivaei}, {Price} and {de
  Groot}}]{Sanders:2015}
{Sanders}, R.~L., {Shapley}, A.~E., {Kriek}, M., {Reddy}, N.~A., {Freeman},
  W.~R., {Coil}, A.~L., {Siana}, B., {Mobasher}, B., {Shivaei}, I., {Price},
  S.~H., {de Groot}, L.: {The MOSDEF Survey: Mass, Metallicity, and
  Star-formation Rate at z $\sim${} 2.3}. \apj \textbf{799},  138 (2015)

\bibitem[{{Scalo} and {Elmegreen}(2004)}]{Scalo:2004}
{Scalo}, J., {Elmegreen}, B.~G.: {Interstellar Turbulence II: Implications and
  Effects}. \araa \textbf{42},  275--316 (2004)

\bibitem[{{Scarano} and {L{\'e}pine}(2013)}]{Scarano:2013}
{Scarano}, S., {L{\'e}pine}, J.~R.~D.: {Radial metallicity distribution breaks
  at corotation radius in spiral galaxies}. \mnras \textbf{428},  625--640
  (2013)

\bibitem[{{Searle}(1971)}]{Searle:1971}
{Searle}, L.: {Evidence for Composition Gradients across the Disks of Spiral
  Galaxies}. \apj \textbf{168},  327 (1971)

\bibitem[{{Sellwood} and {Binney}(2002)}]{Sellwood:2002}
{Sellwood}, J.~A., {Binney}, J.~J.: {Radial mixing in galactic discs}. \mnras
  \textbf{336},  785--796 (2002)

\bibitem[{{Shields}(1974)}]{Shields:1974}
{Shields}, G.~A.: {Composition gradients across spiral galaxies}. \apj
  \textbf{193},  335--341 (1974)

\bibitem[{{Stanghellini} et~al.(2014){Stanghellini}, {Magrini}, {Casasola} and
  {Villaver}}]{Stanghellini:2014}
{Stanghellini}, L., {Magrini}, L., {Casasola}, V., {Villaver}, E.: {The radial
  metallicity gradient and the history of elemental enrichment in M 81 through
  emission-line probes}. \aap \textbf{567},  A88 (2014)

\bibitem[{{Stasi{\'n}ska} et~al.(2012){Stasi{\'n}ska}, {Prantzos}, {Meynet},
  {Sim{\'o}n-D{\'{\i}}az}, {Chiappini}, {Dessauges-Zavadsky}, {Charbonnel},
  {Ludwig}, {Mendoza}, {Grevesse}, {Arnould}, {Barbuy}, {Lebreton},
  {Decourchelle}, {Hill}, {Ferrando}, {H{\'e}brard}, {Durret}, {Katsuma} and
  {Zeippen}}]{Stasinska:2012}
{Stasi{\'n}ska}, G., {Prantzos}, N., {Meynet}, G., {Sim{\'o}n-D{\'{\i}}az}, S.,
  {Chiappini}, C., {Dessauges-Zavadsky}, M., {Charbonnel}, C., {Ludwig}, H.-G.,
  {Mendoza}, C., {Grevesse}, N., {Arnould}, M., {Barbuy}, B., {Lebreton}, Y.,
  {Decourchelle}, A., {Hill}, V., {Ferrando}, P., {H{\'e}brard}, G., {Durret},
  F., {Katsuma}, M., {Zeippen}, C.~J., eds.: {Oxygen in the Universe},
  volume~54 of EAS Publications Series (2012)

\bibitem[{{Thilker} et~al.(2005){Thilker}, {Bianchi}, {Boissier}, {Gil de Paz},
  {Madore}, {Martin}, {Meurer}, {Neff}, {Rich}, {Schiminovich}, {Seibert},
  {Wyder}, {Barlow}, {Byun}, {Donas}, {Forster}, {Friedman}, {Heckman},
  {Jelinsky}, {Lee}, {Malina}, {Milliard}, {Morrissey}, {Siegmund}, {Small},
  {Szalay} and {Welsh}}]{Thilker:2005}
{Thilker}, D.~A., {Bianchi}, L., {Boissier}, S., {Gil de Paz}, A., {Madore},
  B.~F., {Martin}, D.~C., {Meurer}, G.~R., {Neff}, S.~G., {Rich}, R.~M.,
  {Schiminovich}, D., {Seibert}, M., {Wyder}, T.~K., {Barlow}, T.~A., {Byun},
  Y.-I., {Donas}, J., {Forster}, K., {Friedman}, P.~G., {Heckman}, T.~M.,
  {Jelinsky}, P.~N., {Lee}, Y.-W., {Malina}, R.~F., {Milliard}, B.,
  {Morrissey}, P., {Siegmund}, O.~H.~W., {Small}, T., {Szalay}, A.~S., {Welsh},
  B.~Y.: {Recent Star Formation in the Extreme Outer Disk of M83}. \apjl
  \textbf{619},  L79--L82 (2005)

\bibitem[{{Thilker} et~al.(2007){Thilker}, {Bianchi}, {Meurer}, {Gil de Paz},
  {Boissier}, {Madore}, {Boselli}, {Ferguson}, {Mu{\~n}oz-Mateos}, {Madsen},
  {Hameed}, {Overzier}, {Forster}, {Friedman}, {Martin}, {Morrissey}, {Neff},
  {Schiminovich}, {Seibert}, {Small}, {Wyder}, {Donas}, {Heckman}, {Lee},
  {Milliard}, {Rich}, {Szalay}, {Welsh} and {Yi}}]{Thilker:2007}
{Thilker}, D.~A., {Bianchi}, L., {Meurer}, G., {Gil de Paz}, A., {Boissier},
  S., {Madore}, B.~F., {Boselli}, A., {Ferguson}, A.~M.~N., {Mu{\~n}oz-Mateos},
  J.~C., {Madsen}, G.~J., {Hameed}, S., {Overzier}, R.~A., {Forster}, K.,
  {Friedman}, P.~G., {Martin}, D.~C., {Morrissey}, P., {Neff}, S.~G.,
  {Schiminovich}, D., {Seibert}, M., {Small}, T., {Wyder}, T.~K., {Donas}, J.,
  {Heckman}, T.~M., {Lee}, Y.-W., {Milliard}, B., {Rich}, R.~M., {Szalay},
  A.~S., {Welsh}, B.~Y., {Yi}, S.~K.: {A Search for Extended Ultraviolet Disk
  (XUV-Disk) Galaxies in the Local Universe}. \apjs \textbf{173},  538--571
  (2007)

\bibitem[{{Thon} and {Meusinger}(1998)}]{Thon:1998}
{Thon}, R., {Meusinger}, H.: {Models of the long-term evolution of the Galactic
  disk with viscous flows and gas infall}. \aap \textbf{338},  413--434 (1998)

\bibitem[{{Toribio San Cipriano} et~al.(2016){Toribio San Cipriano},
  {Garc{\'{\i}}a-Rojas}, {Esteban}, {Bresolin} and
  {Peimbert}}]{Toribio-San-Cipriano:2016}
{Toribio San Cipriano}, L., {Garc{\'{\i}}a-Rojas}, J., {Esteban}, C.,
  {Bresolin}, F., {Peimbert}, M.: {Carbon and oxygen abundance gradients in NGC
  300 and M33 from optical recombination lines}. \mnras \textbf{458},
  1866--1890 (2016)

\bibitem[{{Torres-Flores} et~al.(2014){Torres-Flores}, {Scarano}, {Mendes de
  Oliveira}, {de Mello}, {Amram} and {Plana}}]{Torres-Flores:2014}
{Torres-Flores}, S., {Scarano}, S., {Mendes de Oliveira}, C., {de Mello},
  D.~F., {Amram}, P., {Plana}, H.: {Star-forming regions and the metallicity
  gradients in the tidal tails: the case of NGC 92}. \mnras \textbf{438},
  1894--1908 (2014)

\bibitem[{{Torrey} et~al.(2012){Torrey}, {Cox}, {Kewley} and
  {Hernquist}}]{Torrey:2012}
{Torrey}, P., {Cox}, T.~J., {Kewley}, L., {Hernquist}, L.: {The Metallicity
  Evolution of Interacting Galaxies}. \apj \textbf{746},  108 (2012)

\bibitem[{{Tosi}(1988)}]{Tosi:1988}
{Tosi}, M.: {The effect of metal-rich infall on galactic chemical evolution}.
  \aap \textbf{197},  47--51 (1988)

\bibitem[{{Tremonti} et~al.(2004){Tremonti}, {Heckman}, {Kauffmann},
  {Brinchmann}, {Charlot}, {White}, {Seibert}, {Peng}, {Schlegel}, {Uomoto},
  {Fukugita} and {Brinkmann}}]{Tremonti:2004}
{Tremonti}, C.~A., {Heckman}, T.~M., {Kauffmann}, G., {Brinchmann}, J.,
  {Charlot}, S., {White}, S.~D.~M., {Seibert}, M., {Peng}, E.~W., {Schlegel},
  D.~J., {Uomoto}, A., {Fukugita}, M., {Brinkmann}, J.: {The Origin of the
  Mass-Metallicity Relation: Insights from 53,000 Star-forming Galaxies in the
  Sloan Digital Sky Survey}. \apj \textbf{613},  898--913 (2004)

\bibitem[{{Tsujimoto} et~al.(1995){Tsujimoto}, {Yoshii}, {Nomoto} and
  {Shigeyama}}]{Tsujimoto:1995}
{Tsujimoto}, T., {Yoshii}, Y., {Nomoto}, K., {Shigeyama}, T.: {Abundance
  gradients in the star-forming viscous disk and chemical properties of the
  bulge.} \aap \textbf{302},  704 (1995)

\bibitem[{{Tumlinson} et~al.(2011)}]{Tumlinson:2011}
{Tumlinson}, J., et~al.: {The Large, Oxygen-Rich Halos of Star-Forming Galaxies
  Are a Major Reservoir of Galactic Metals}. Science \textbf{334},  948 (2011)

\bibitem[{{Vale Asari} et~al.(2016){Vale Asari}, {Stasi{\'n}ska}, {Morisset}
  and {Cid Fernandes}}]{Vale-Asari:2016}
{Vale Asari}, N., {Stasi{\'n}ska}, G., {Morisset}, C., {Cid Fernandes}, R.:
  {BOND: Bayesian Oxygen and Nitrogen abundance Determinations in giant H II
  regions using strong and semi-strong lines}. \mnras  (2016)

\bibitem[{{van Zee} et~al.(1998){van Zee}, {Salzer}, {Haynes}, {O'Donoghue} and
  {Balonek}}]{van-Zee:1998a}
{van Zee}, L., {Salzer}, J.~J., {Haynes}, M.~P., {O'Donoghue}, A.~A.,
  {Balonek}, T.~J.: {Spectroscopy of Outlying H II Regions in Spiral Galaxies:
  Abundances and Radial Gradients}. \aj \textbf{116},  2805--2833 (1998)

\bibitem[{{Vila-Costas} and {Edmunds}(1992)}]{Vila-Costas:1992}
{Vila-Costas}, M.~B., {Edmunds}, M.~G.: {The relation between abundance
  gradients and the physical properties of spiral galaxies}. \mnras
  \textbf{259},  121--145 (1992)

\bibitem[{{Vila Costas} and {Edmunds}(1993)}]{Vila-Costas:1993}
{Vila Costas}, M.~B., {Edmunds}, M.~G.: {The Nitrogen-To Ratio in Galaxies and
  its Implications for the Origin of Nitrogen}. \mnras \textbf{265},  199--212
  (1993)

\bibitem[{{V{\'{\i}}lchez} and {Esteban}(1996)}]{Vilchez:1996}
{V{\'{\i}}lchez}, J.~M., {Esteban}, C.: {The chemical composition of HII
  regions in the outer Galaxy}. \mnras \textbf{280},  720--734 (1996)

\bibitem[{{Vlaji{\'c}}, {Bland-Hawthorn} and {Freeman}(2009)}]{Vlajic:2009}
{Vlaji{\'c}}, M., {Bland-Hawthorn}, J., {Freeman}, K.~C.: {The Abundance
  Gradient in the Extremely Faint Outer Disk of NGC 300}. \apj \textbf{697},
  361--372 (2009)

\bibitem[{{Vlaji{\'c}}, {Bland-Hawthorn} and {Freeman}(2011)}]{Vlajic:2011}
{Vlaji{\'c}}, M., {Bland-Hawthorn}, J., {Freeman}, K.~C.: {The Structure and
  Metallicity Gradient in the Extreme Outer Disk of NGC 7793}. \apj
  \textbf{732},  7 (2011)

\bibitem[{{Wang} et~al.(2013){Wang}, {Kauffmann}, {J{\'o}zsa}, {Serra}, {van
  der Hulst}, {Bigiel}, {Brinchmann}, {Verheijen}, {Oosterloo}, {Wang}, {Li},
  {den Heijer} and {Kerp}}]{Wang:2013}
{Wang}, J., {Kauffmann}, G., {J{\'o}zsa}, G.~I.~G., {Serra}, P., {van der
  Hulst}, T., {Bigiel}, F., {Brinchmann}, J., {Verheijen}, M.~A.~W.,
  {Oosterloo}, T., {Wang}, E., {Li}, C., {den Heijer}, M., {Kerp}, J.: {The
  Bluedisks project, a study of unusually H I-rich galaxies - I. H I sizes and
  morphology}. \mnras \textbf{433},  270--294 (2013)

\bibitem[{{Werk} et~al.(2013){Werk}, {Prochaska}, {Thom}, {Tumlinson}, {Tripp},
  {O'Meara} and {Peeples}}]{Werk:2013}
{Werk}, J.~K., {Prochaska}, J.~X., {Thom}, C., {Tumlinson}, J., {Tripp}, T.~M.,
  {O'Meara}, J.~M., {Peeples}, M.~S.: {The COS-Halos Survey: An Empirical
  Description of Metal-line Absorption in the Low-redshift Circumgalactic
  Medium}. \apjs \textbf{204},  17 (2013)

\bibitem[{{Werk} et~al.(2010){Werk}, {Putman}, {Meurer}, {Ryan-Weber},
  {Kehrig}, {Thilker}, {Bland-Hawthorn}, {Drinkwater}, {Kennicutt}, {Wong},
  {Freeman}, {Oey}, {Dopita}, {Doyle}, {Ferguson}, {Hanish}, {Heckman},
  {Kilborn}, {Kim}, {Knezek}, {Koribalski}, {Meyer}, {Smith} and
  {Zwaan}}]{Werk:2010}
{Werk}, J.~K., {Putman}, M.~E., {Meurer}, G.~R., {Ryan-Weber}, E.~V., {Kehrig},
  C., {Thilker}, D.~A., {Bland-Hawthorn}, J., {Drinkwater}, M.~J., {Kennicutt},
  R.~C., {Wong}, O.~I., {Freeman}, K.~C., {Oey}, M.~S., {Dopita}, M.~A.,
  {Doyle}, M.~T., {Ferguson}, H.~C., {Hanish}, D.~J., {Heckman}, T.~M.,
  {Kilborn}, V.~A., {Kim}, J.~H., {Knezek}, P.~M., {Koribalski}, B., {Meyer},
  M., {Smith}, R.~C., {Zwaan}, M.~A.: {Outlying H II Regions in H I-Selected
  Galaxies}. \aj \textbf{139},  279--295 (2010)

\bibitem[{{Werk} et~al.(2011){Werk}, {Putman}, {Meurer} and
  {Santiago-Figueroa}}]{Werk:2011}
{Werk}, J.~K., {Putman}, M.~E., {Meurer}, G.~R., {Santiago-Figueroa}, N.:
  {Metal Transport to the Gaseous Outskirts of Galaxies}. \apj \textbf{735},
  71 (2011)

\bibitem[{{Worthey} et~al.(2005){Worthey}, {Espa{\~n}a}, {MacArthur} and
  {Courteau}}]{Worthey:2005}
{Worthey}, G., {Espa{\~n}a}, A., {MacArthur}, L.~A., {Courteau}, S.: {M31's
  Heavy-Element Distribution and Outer Disk}. \apj \textbf{631},  820--831
  (2005)

\bibitem[{{Wyder} et~al.(2009){Wyder}, {Martin}, {Barlow}, {Foster},
  {Friedman}, {Morrissey}, {Neff}, {Neill}, {Schiminovich}, {Seibert},
  {Bianchi}, {Donas}, {Heckman}, {Lee}, {Madore}, {Milliard}, {Rich}, {Szalay}
  and {Yi}}]{Wyder:2009}
{Wyder}, T.~K., {Martin}, D.~C., {Barlow}, T.~A., {Foster}, K., {Friedman},
  P.~G., {Morrissey}, P., {Neff}, S.~G., {Neill}, J.~D., {Schiminovich}, D.,
  {Seibert}, M., {Bianchi}, L., {Donas}, J., {Heckman}, T.~M., {Lee}, Y.,
  {Madore}, B.~F., {Milliard}, B., {Rich}, R.~M., {Szalay}, A.~S., {Yi}, S.~K.:
  {The Star Formation Law at Low Surface Density}. \apj \textbf{696},
  1834--1853 (2009)

\bibitem[{{Yang} and {Krumholz}(2012)}]{Yang:2012}
{Yang}, C.-C., {Krumholz}, M.: {Thermal-instability-driven Turbulent Mixing in
  Galactic Disks. I. Effective Mixing of Metals}. \apj \textbf{758},  48 (2012)

\bibitem[{{Yong}, {Carney} and {Friel}(2012)}]{Yong:2012}
{Yong}, D., {Carney}, B.~W., {Friel}, E.~D.: {Elemental Abundance Ratios in
  Stars of the Outer Galactic Disk. IV. A New Sample of Open Clusters}. \aj
  \textbf{144},  95 (2012)

\bibitem[{{Zahid} and {Bresolin}(2011)}]{Zahid:2011a}
{Zahid}, H.~J., {Bresolin}, F.: {Reexamination of the Radial Abundance Gradient
  Break in NGC 3359}. \aj \textbf{141},  192 (2011)

\bibitem[{{Zahid} et~al.(2013){Zahid}, {Geller}, {Kewley}, {Hwang}, {Fabricant}
  and {Kurtz}}]{Zahid:2013}
{Zahid}, H.~J., {Geller}, M.~J., {Kewley}, L.~J., {Hwang}, H.~S., {Fabricant},
  D.~G., {Kurtz}, M.~J.: {The Chemical Evolution of Star-forming Galaxies over
  the Last 11 Billion Years}. \apjl \textbf{771},  L19 (2013)

\bibitem[{{Zaritsky}, {Kennicutt} and {Huchra}(1994)}]{Zaritsky:1994}
{Zaritsky}, D., {Kennicutt}, R.~C., Jr., {Huchra}, J.~P.: {H II regions and the
  abundance properties of spiral galaxies}. \apj \textbf{420},  87--109 (1994)

\bibitem[{{Zinchenko} et~al.(2015){Zinchenko}, {Berczik}, {Grebel}, {Pilyugin}
  and {Just}}]{Zinchenko:2015}
{Zinchenko}, I.~A., {Berczik}, P., {Grebel}, E.~K., {Pilyugin}, L.~S., {Just},
  A.: {On the Influence of Minor Mergers on the Radial Abundance Gradient in
  Disks of Milky-Way-like Galaxies}. \apj \textbf{806},  267 (2015)

\bibitem[{{Zurita} and {Bresolin}(2012)}]{Zurita:2012}
{Zurita}, A., {Bresolin}, F.: {The chemical abundance in M31 from H II
  regions}. \mnras \textbf{427},  1463--1481 (2012)

\end{thebibliography}



\end{document}